\documentclass
[aps,pra,amsfonts,amssymb,twocolumn,amsmath,preprintnumbers,nofootinbib,floatfix,
showpacs]{revtex4-1}%
\usepackage[dvips]{graphics}
\usepackage{graphicx}
\usepackage{bm}
\usepackage{amsmath}
\usepackage{amsfonts}
\usepackage{amssymb}
\usepackage{xcolor}
\usepackage{subfigure}
\usepackage{hyperref,hypcap}
\usepackage{braket}
\usepackage{commath}%
\providecommand{\U}[1]{\protect\rule{.1in}{.1in}}

\begin{document}

\title{Boltzmann approach to spin-orbit-induced transport in effective quantum theories}
\author{Cong Xiao, Bangguo Xiong, Fei Xue}
\affiliation{Department of Physics, The University of Texas, Austin, Texas 78712-0264, USA}

\begin{abstract}
In model studies of the spin/anomalous Hall effect, effective Hamiltonians
often serve as the starting point. However, a complete effective quantum
theory contains not only the effective Hamiltonian but also the relation
linking the physical observables to the canonical ones. We construct the
semiclassical Boltzmann (SB) transport framework in the weak
disorder-potential regime directly in the level of the effective quantum
theory, and confirm this construction by formulating a generalized
Kohn-Luttinger density matrix transport theory also in this level. The link
and difference between the present SB theory and previous phenomenological
Boltzmann, quantum kinetic and usual Kubo-Streda theories are clarified. We
also present the slightly generalized Kubo-Streda formula in the level of the
effective quantum theory. In this level, it is this generalized Kubo-Streda
formula rather than the usual one that leads to the same physical
interpretations as the present SB theory. In the application to a Rashba 2D
effective model, a nonzero spin Hall effect important in the case of strong
Rashba coupling but neglected in previous theories is found.
\end{abstract}
\maketitle

\section{Introduction}

In model studies of the spin/anomalous Hall effect -- a spin-orbit-induced
transverse spin/charge transport effect \cite{Nagaosa2010,Sinova2015},
effective Hamiltonians, such as the two-band Rashba model \cite{Sinova2004},
often serve as the starting point. The spin-orbit interaction (SOI) appearing
in the external-perturbation-free effective Hamiltonian is often termed as the
band-structure SOI. The effective Hamiltonian is only part of the effective
quantum theory, thus may not be enough for predictions relevant to experiments
\cite{Niu2010,Chang2008}. A complete effective quantum theory contains not
only the effective Hamiltonian in the presence of external perturbations but
also the relation linking the physical observables to the canonical ones. This
is in fact clear in the seminal work of Nozieres and Lewiner \cite{1973}, who
obtained the effective quantum theory for conduction electrons from the parent
eight-band Kane model in direct-gap III-V semiconductors. The physical
position operator in this effective quantum theory differs from the canonical
one by an additional term arising from the projection from the eight-band to
the two-band models \cite{Niu2010,Chang2008,Engel2005}. This additional term
gives rise to sizable effective SOIs with the external electric field and with
impurities in the level of the effective quantum theory \cite{1973}.

Unfortunately, the complicated and phenomenological Boltzmann treatment of
Nozieres and Lewiner on the Hall transport did not yield a simple picture for
the many terms they obtained. Perhaps partly due to this reason, many
subsequent theories using the phenomenological-Boltzmann
\cite{Lyo1972,Zhang2000} or usual Kubo-Streda (zero-frequency linear response
\cite{Streda2010}) diagrammatic \cite{Bruno2001,Dugaev2001,Tse2006PRL}
approaches still considered only the effective Hamiltonian without SOI with
the driving external electric field and neglected the change of the physical
position. Although some recent phenomenological-Boltzmann
\cite{Vignale2006PRB,Vignale2009,Vignale2009JPCM} and quantum kinetic theories
\cite{Culcer2010} considered the complete Nozieres-Lewiner effective model,
the link and difference between various theories applied to this simple model
have not yet been completely clarified. Moreover, as we will reveal, these
different considerations may lead to different spin Hall conductivities when
the external-perturbation-free effective Hamiltonian has its own internal
structure, i.e., the band-structure SOI in the level of the effective theory.
Therefore, a simple transport-theory framework with physical insights in the
level of the effective quantum theory is highly desirable.

The semiclassical Boltzmann (SB) transport theory is appealing because it is
conceptually simple \cite{Ziman1960,Sinitsyn2008,Nagaosa2010} and has the
microscopic density matrix approach as its solid foundation
\cite{KL1957,Luttinger1958,Sinitsyn2006,Xiao2017SOT-SBE,Xiao2018scaling}. When
the SB and equivalent theories \cite{Sinitsyn2007,Xiao2018scaling} apply, the
spin/anomalous Hall effect can be parsed clearly in the presence of static
disorder. Three mechanisms -- intrinsic, anomalous quantum (called side-jump
in recent reviews \cite{Nagaosa2010,Sinova2015}) and skew scattering -- are
defined unambiguously \cite{Nagaosa2010,Sinova2015}. For the spin/anomalous
Hall conductivity, the intrinsic contribution is independent of disorder, the
anomalous quantum contribution relies on the disorder but turns out to be
independent of the impurity density, whereas the skew scattering contribution
from disorder is inversely proportional to the impurity density. However,
existing SB theory and the underlying Kohn-Luttinger density matrix theory are
only formulated in the level of the full Hamiltonian where the physical
observables are just the canonical ones \cite{Niu2010,Chang2008}.

In this paper we construct the SB framework in the weak disorder-potential
regime directly in the level of the effective quantum theory, and confirm this
construction by formulating a generalized Kohn-Luttinger theory also in this
level \cite{note-SB}. It is shown that the spin/anomalous Hall effect studied
in this level can still be parsed into the same categories as in the level of
the full Hamiltonian, in the regime where the SB theory works
\cite{Xiao2018scaling}. We discuss the link and difference between the present
SB theory and previous phenomenological Boltzmann
\cite{Lyo1972,Zhang2000,Vignale2006PRB,Vignale2009,Vignale2009JPCM}, quantum
kinetic \cite{Raimondi2012,Culcer2013,Shen2014} and usual Kubo-Streda
diagrammatic \cite{Bruno2001,Dugaev2001,Tse2006PRL,Tse2006PRB} theories. To
help clarify this issue, we also derive the slightly generalized Kubo-Streda
formula in the level of the effective quantum theory. In this level, it is
this generalized Kubo-Streda formula rather than the usual one
\cite{Bruno2001,Streda2010} that leads to the same physical picture as the
present SB theory.

The SB picture is valid in the Boltzmann regime where the disorder-broadening
of bands is quite smaller than the minimal intrinsic energy-scale around the
Fermi level \cite{Xiao2018scaling}. It is easy to reach this regime even in
moderately dirty systems if the minimal intrinsic energy-scale around the
Fermi level is quite large. However, in the opposite case, the system may be
located within the so-called diffusive regime (limit) \cite{Xiao2017SOT} where
a drift-diffusion equation for coupled spin-charge dynamics holds
\cite{Raimondi2012,Shen2014}. Both the SB and drift-diffusion theories can
only work in limited and different regimes \cite{note-regime} with different
physical pictures.

The paper is organized as follows. The SB theory is formulated in Sec. II and
III, whereas further model-analysis and comparison with other theories are
presented in Sec. IV and V. Before summarizing the paper in Sec. VI we discuss
the validity of a widely-accepted idea proposed in a previous phenomenological
Boltzmann theory. The generalized Kohn-Luttinger and Kubo-Streda approaches in
the level of the effective quantum theory are presented in Appendix A and C,
respectively. Appendix B and D contains some other details supporting the
discussion in the maintext.

\section{Effective quantum theory}

The total single-carrier effective Hamiltonian reads \cite{1973}
\begin{equation}
\hat{H}_{T}=\hat{H}_{0}+\hat{V}\left(  \mathbf{\hat{r}}^{phy}\right)
-e\mathbf{E\cdot\hat{r}}^{phy}, \label{model}%
\end{equation}
where the physical position operator $\mathbf{\hat{r}}^{phy}$ may differ from
the canonical one $\mathbf{\hat{r}}$. Hereafter $\hat{V}\left(  \mathbf{\hat
{r}}^{phy}\right)  $ and $\hat{V}\left(  \mathbf{\hat{r}}\right)  $ are
sometimes designated as $\hat{W}$ and $\hat{V}$, respectively, for brevity.
The velocity operator in the presence of disorder and external electric field
is $\mathbf{\hat{v}}=\frac{1}{i\hbar}\left[  \mathbf{\hat{r}}^{phy},\hat
{H}_{T}\right]  =\mathbf{\hat{v}}^{phy}+\delta^{V}\mathbf{\hat{v}}%
+\delta^{\mathbf{E}}\mathbf{\hat{v}}$, where $\mathbf{\hat{v}}^{phy}%
\equiv\frac{1}{i\hbar}\left[  \mathbf{\hat{r}}^{phy},\hat{H}_{0}\right]  $,
$\delta^{V}\mathbf{\hat{v}}\equiv\frac{1}{i\hbar}\left[  \mathbf{\hat{r}%
}^{phy},\hat{V}\left(  \mathbf{\hat{r}}^{phy}\right)  \right]  $\ and
$\delta^{\mathbf{E}}\mathbf{\hat{v}}\equiv\frac{1}{i\hbar}\left[
\mathbf{\hat{r}}^{phy},-e\mathbf{E\cdot\hat{r}}^{phy}\right]  $. For 2D
electrons or holes \cite{Culcer2013},%
\begin{equation}
\mathbf{\hat{r}}^{phy}=\mathbf{\hat{r}+\hat{r}}^{a}=\mathbf{\hat{r}+}%
\frac{\lambda_{0}^{2}}{4}\mathbf{\hat{\sigma}\times\hat{K},}%
\end{equation}
where $\mathbf{\hat{K}}=\left(  \hat{k}_{x}^{3}-3\hat{k}_{y}^{2}\hat{k}%
_{x},3\hat{k}_{x}^{2}\hat{k}_{y}-\hat{k}_{y}^{3},0\right)  $ for 2D holes and
$\mathbf{\hat{K}=\hat{k}}$ for 2D electrons, $\mathbf{\hat{k}}=\left(  \hat
{k}_{x},\hat{k}_{y}\right)  $ is the momentum operator. In the band-eigenstate
representation of the disorder-free effective Hamiltonian $\hat{H}_{0}$,
\begin{equation}
\left(  \mathbf{\hat{r}}^{phy}\right)  _{ll^{\prime}}=i\partial_{\mathbf{k}%
}\delta_{ll^{\prime}}+i\mathbf{\tilde{J}}_{ll^{\prime}},
\end{equation}
with $i\mathbf{\tilde{J}}_{ll^{\prime}}\equiv i\mathbf{J}_{ll^{\prime}%
}+i\mathbf{J}_{ll^{\prime}}^{a}$, $\mathbf{J}_{ll^{\prime}}=\delta
_{\mathbf{kk}^{\prime}}\langle u_{l}|\partial_{\mathbf{k}}|u_{l^{\prime}%
}\rangle$ and $i\mathbf{J}_{ll^{\prime}}^{a}=\delta_{\mathbf{kk}^{\prime}%
}\frac{\lambda_{0}^{2}}{4}\mathbf{\hat{\sigma}}_{ll^{\prime}}\times\mathbf{K}%
$. $\mathbf{|}l\rangle\equiv|\eta\mathbf{k}\rangle=|\mathbf{k}\rangle
|u_{\mathbf{k}}^{\eta}\rangle$ is the eigenstate of $\hat{H}_{0}$\ with energy
$\epsilon_{l}\equiv\epsilon_{\mathbf{k}}^{\eta}$, $\eta$ the band index.

\section{SB framework in the weak disorder-potential regime}

Regarding the linear response to a constant electric field $\mathbf{E}$ in
non-degenerate multiband electron systems with weak static disorder, in the SB
framework the average value of physical quantity $A$ is expressed as%
\begin{equation}
A=\sum_{l}A_{l}f_{l}. \label{Boltzmann}%
\end{equation}
The semiclassical distribution function $f_{l}=f_{l}^{0}+g_{l}+g_{l}^{a}$ is
expanded up to the linear order of $\mathbf{E}$\ around the equilibrium Fermi
distribution function $f^{0}$, $g_{l}$ and $g_{l}^{a}$ equilibrate the effects
of the driving electric field between and during successive scattering events,
respectively \cite{Sinitsyn2005,Sinitsyn2006,Sinitsyn2008}. These processes
are accounted for by the linearized SB equation in nonequilibrium
steady-states \cite{Sinitsyn2007,Sinitsyn2008}%
\begin{gather}
e\mathbf{E}\cdot\mathbf{v}_{l}^{0}\frac{\partial f^{0}}{\partial\epsilon_{l}%
}=-\sum_{l^{\prime}}\omega_{ll^{\prime}}\left(  g_{l}-g_{l^{\prime}}\right)
,\label{SBE-n}\\
0=\sum_{l^{\prime}}\omega_{ll^{\prime}}^{\left(  2\right)  }\left(  g_{l}%
^{a}-g_{l^{\prime}}^{a}-\frac{\partial f^{0}}{\partial\epsilon_{l}}%
e\mathbf{E}\cdot\delta\mathbf{\tilde{r}}_{l^{\prime}l}\right)  . \label{SBE-a}%
\end{gather}
Here $\mathbf{v}_{l}^{0}=\partial\epsilon_{l}/\hbar\partial\mathbf{k}$ is the
group velocity, $\omega_{ll^{\prime}}=\frac{2\pi}{\hbar}\left\langle
\left\vert T_{ll^{\prime}}\right\vert ^{2}\right\rangle \delta\left(
d_{ll^{\prime}}\right)  $ is the scattering rate $\left(  l^{\prime
}\rightarrow l\right)  $, $\left\langle ..\right\rangle $ denotes the disorder
average. Hereafter we employ the simplified notations $d_{ll^{\prime}}%
\equiv\epsilon_{l}-\epsilon_{l^{\prime}}$ and $d_{ll^{\prime}}^{\pm}\equiv
d_{ll^{\prime}}\pm i\hbar s$. $s\rightarrow0^{+}$ appears to be the
regularizing factor in the T-matrix theory.\ The T-matrix is determined by the
Lippmann-Schwinger equation $\hat{T}=\hat{W}+\hat{W}\left(  \epsilon_{l}%
-\hat{H}_{0}+i\hbar s\right)  ^{-1}\hat{T}$. In the weak disorder-potential
regime $\omega_{ll^{\prime}}$ is obtained by expanding $\left\langle
\left\vert T_{ll^{\prime}}\right\vert ^{2}\right\rangle $ in terms of the
disorder-potential: $\omega_{ll^{\prime}}=\omega_{ll^{\prime}}^{\left(
2\right)  }+\omega_{ll^{\prime}}^{3a}+\omega_{ll^{\prime}}^{4a}$. The number
and \textquotedblleft a\textquotedblright\ in the superscripts mean the order
of disorder potential and the anti-symmetric part ($\omega_{ll^{\prime}}%
^{a}=\frac{1}{2}\left(  \omega_{ll^{\prime}}-\omega_{l^{\prime}l}\right)  $)
of the scattering rate, respectively. In the weak disorder-potential regime
the expansion up to the third Born order is sufficient. Taking into account
the symmetric part of the higher-order scattering rate only renormalizes the
longitudinal distribution function \cite{Sinitsyn2008,Nagaosa2010} and is not
necessary in the Boltzmann regime in the case of weak disorder potential.
Therefore, $g_{l}=g_{l}^{\left(  -2\right)  }+g_{l}^{\left(  -1\right)
}+g_{l}^{\left(  0\right)  }$ and concurrently we can set $g_{l}^{\left(
-1\right)  }=g_{l}^{sk}$ and $g_{l}^{\left(  0\right)  }=g_{l}^{pair}$ to
emphasize the physic related to these distribution functions: $g_{l}^{sk}$
arises from the skew scattering and $g_{l}^{pair}$ from scattering off pairs
of impurities (details in Refs. \cite{Sinitsyn2008,Xiao2018scaling}).

For the coordinate-shift \cite{Sinitsyn2008} $\delta\mathbf{\tilde{r}%
}_{l^{\prime}l}$, the wavepacket derivation of Sinitsyn et al.
\cite{Sinitsyn2006} who assumed $\mathbf{\hat{r}}=\mathbf{\hat{r}}^{phy}$ can
be directly generalized to the present case ($\mathbf{\hat{r}}\rightarrow
\mathbf{\hat{r}}^{phy}$, $\hat{V}\left(  \mathbf{\hat{r}}\right)
\rightarrow\hat{V}\left(  \mathbf{\hat{r}}^{phy}\right)  =\hat{W}$), yielding%
\begin{equation}
\delta\mathbf{\tilde{r}}_{l^{\prime}l}=i\mathbf{\tilde{J}}_{l^{\prime}%
}-i\mathbf{\tilde{J}}_{l}-\mathbf{\hat{D}}\arg W_{l^{\prime}l},
\end{equation}
with $i\mathbf{\tilde{J}}_{l}\equiv i\mathbf{\tilde{J}}_{ll}$ and
$\mathbf{\hat{D}}=\partial_{\mathbf{k}^{\prime}}+\partial_{\mathbf{k}}$.

In Eq. (\ref{Boltzmann}), an appropriate expression for $A_{l}$ is needed. In
the conventional SB consideration one takes $A_{l}=\langle l|\left(  \hat
{A}^{phy}+\delta^{V}\hat{A}+\delta^{\mathbf{E}}\hat{A}\right)  |l\rangle$. For
spin-orbit-induced transport, the electric-field-induced and impurity-induced
corrections to $|l\rangle$ contribute qualitatively \cite{Xiao2017SOT-SBE},
thus $|l\rangle\rightarrow|l\rangle+|\delta^{\mathbf{E}}l\rangle+|\delta
^{V}l\rangle$. Here $|\delta^{\mathbf{E}}l\rangle=-e\mathbf{E\cdot}%
\sum_{l^{\prime}\neq l}|l^{\prime}\rangle\langle l^{\prime}|\mathbf{\hat{r}%
}^{phy}|l\rangle/d_{ll^{\prime}}^{+}$ is the electric-field induced correction
to the Bloch state, whereas $|\delta^{V}l\rangle=\left(  \epsilon_{l}-\hat
{H}_{0}+i\hbar s\right)  ^{-1}\hat{T}|l\rangle$ is the scattering correction.
Therefore we have
\begin{equation}
A_{l}\equiv\tilde{A}_{l}^{0}+\delta^{in}A_{l}+\delta^{in,1}A_{l}+\delta
^{sj}A_{l}+\delta^{sj,1}A_{l},
\end{equation}
where $\tilde{A}_{l}^{0}=\langle l|\hat{A}^{phy}|l\rangle$ and
\begin{equation}
\delta^{in}A_{l}=2\operatorname{Re}\langle l|\hat{A}^{phy}|\delta^{\mathbf{E}%
}l\rangle,\text{ }\delta^{in,1}A_{l}=\langle l|\delta^{\mathbf{E}}\hat
{A}|l\rangle,
\end{equation}
and%
\begin{gather}
\delta^{sj}A_{l}=2\operatorname{Re}\left\langle \langle l|\hat{A}^{phy}%
|\delta^{V}l\rangle\right\rangle ^{off}+\left\langle \langle\delta^{V}%
l|\hat{A}^{phy}|\delta^{V}l\rangle\right\rangle ^{off},\nonumber\\
\delta^{sj,1}A_{l}=2\operatorname{Re}\left\langle \langle l|\delta^{V}\hat
{A}|\delta^{V}l\rangle\right\rangle ^{off}.
\end{gather}
The superscript \textquotedblleft$off$\textquotedblright\ means that only the
terms with off-diagonal elements of $\hat{A}^{phy}$ or $\delta^{V}\hat{A}$ in
the Bloch representation $\left\{  |\eta\mathbf{k}\rangle\right\}  $ are
retained. $\delta^{sj}A_{l}$ contains $\eta$-off-diagonal interband matrix
elements of $\hat{A}^{phy}$ \cite{Xiao2017SOT-SBE}, whereas $\delta
^{sj,1}A_{l}$ contains\ $\mathbf{k}$-off-diagonal matrix elements
of\ $\delta^{V}\hat{A}$ \cite{Lyo1972}. In the weak disorder-potential regime,
we only preserve $|\delta^{V}l\rangle$ to the lowest nonzero Born order when
calculating $\delta^{sj}A_{l}$ and $\delta^{sj,1}A_{l}$. More precisely, we
obtain the following expressions%
\begin{equation}
\delta^{in}A_{l}=-\hbar e\mathbf{E}\cdot\sum_{l^{\prime}\neq l}\delta
_{\mathbf{kk}^{\prime}}\frac{2\mathrm{Im}\langle u_{l}|\mathbf{\hat{v}}%
^{phy}|u_{l^{\prime}}\rangle A_{l^{\prime}l}^{phy}}{d_{ll^{\prime}}^{2}},
\end{equation}%
\begin{align}
\delta^{sj}A_{l}  &  =\sum_{l^{\prime},l^{\prime\prime}\neq l^{\prime}}%
\frac{\left\langle W_{ll^{\prime}}W_{l^{\prime\prime}l}\right\rangle
A_{l^{\prime}l^{\prime\prime}}^{phy}}{d_{ll^{\prime}}^{-}d_{ll^{\prime\prime}%
}^{+}}\label{sj}\\
&  +2\operatorname{Re}\sum_{l^{\prime}\neq l,l^{\prime\prime}}\frac
{\left\langle W_{l^{\prime}l^{\prime\prime}}W_{l^{\prime\prime}l}\right\rangle
A_{ll^{\prime}}^{phy}}{d_{ll^{\prime}}^{+}d_{ll^{\prime\prime}}^{+}},\nonumber
\end{align}
and
\begin{equation}
\delta^{sj,1}A_{l}=2\operatorname{Re}\sum_{l^{\prime}\neq l}\frac{\left\langle
\langle l|\delta^{V}\hat{A}\mathbf{|}l^{\prime}\rangle W_{l^{\prime}%
l}\right\rangle }{d_{ll^{\prime}}^{+}}.
\end{equation}

The linear response of $A$ in the weak disorder-potential regime thus reads
\begin{gather}
\delta A=\sum_{l}\left(  g_{l}^{\left(  -2\right)  }+g_{l}^{sk}\right)
\tilde{A}_{l}^{0}+\sum_{l}\left(  g_{l}^{a}+g_{l}^{pair}\right)  \tilde{A}%
_{l}^{0}\label{linear response}\\
+\sum_{l}g_{l}^{\left(  -2\right)  }\left(  \delta^{sj}A_{l}+\delta
^{sj,1}A_{l}\right)  +\sum_{l}f_{l}^{0}\left(  \delta^{in}A_{l}+\delta
^{in,1}A_{l}\right)  .\nonumber
\end{gather}
The last term on the right-hand-side (rhs) is the intrinsic contribution. The
second and third terms constitute the anomalous quantum contribution mentioned
in the introduction, which is also independent of the disorder potential in
the weak disorder-potential regime \cite{Yang2011}.

The SB formalism presented in this section is confirmed by formulating the
Kohn-Luttinger density matrix transport approach designed in the weak disorder
potential regime \cite{KL1957,Luttinger1958} directly in the level of the
effective quantum theory. This microscopic approach is presented in Appendix A.

\subsection{Anomalous Hall effect}

To be more specific, we analyze the anomalous Hall effect. In the level of the
full Hamiltonian $\mathbf{\hat{r}}=\mathbf{\hat{r}}^{phy}$, $\delta
^{sj}\mathbf{v}_{l}$ equals the semiclassical side-jump velocity
$\mathbf{v}_{l}^{sj}=\sum_{l^{\prime}}\omega_{ll^{\prime}}^{\left(  2\right)
}\delta\mathbf{r}_{l^{\prime}l}$%
\ \cite{Sinitsyn2006,Sinitsyn2008,Xiao2017SOT-SBE}, and $\delta^{in}%
\mathbf{v}_{l}$ is just the Berry-curvature anomalous velocity $\mathbf{v}%
_{l}^{B}=\Omega_{l}^{0}\mathbf{\hat{z}}\times\frac{e}{\hbar}\mathbf{E}$. Here
$\delta\mathbf{r}_{l^{\prime}l}=i\mathbf{J}_{l^{\prime}}-i\mathbf{J}%
_{l}-\mathbf{\hat{D}}\arg V_{l^{\prime}l}$ and $\Omega_{l}^{0}=\partial
_{k_{x}}\left(  i\mathbf{J}_{l}\right)  _{y}-\partial_{k_{y}}\left(
i\mathbf{J}_{l}\right)  _{x}$. In the level of the effective theory, we prove
(Appendix B)
\begin{equation}
\delta^{sj}\mathbf{v}_{l}+\delta^{sj,1}\mathbf{v}_{l}=\sum_{l^{\prime}}%
\omega_{ll^{\prime}}^{\left(  2\right)  }\delta\mathbf{\tilde{r}}_{l^{\prime
}l}\equiv\mathbf{\tilde{v}}_{l}^{sj} \label{sj-velocity}%
\end{equation}
and (Appendix B)
\begin{equation}
\delta^{in}\mathbf{v}_{l}+\delta^{in,1}\mathbf{v}_{l}=\left(  \Omega_{l}%
^{0}+\Omega_{l}^{a}\right)  \mathbf{\hat{z}}\times\frac{e}{\hbar}%
\mathbf{E}\equiv\mathbf{\tilde{v}}_{l}^{B}, \label{Berry-velocity}%
\end{equation}
where $\Omega_{l}^{a}\equiv\partial_{k_{x}}\left(  i\mathbf{J}_{l}^{a}\right)
_{y}-\partial_{k_{y}}\left(  i\mathbf{J}_{l}^{a}\right)  _{x}$. Therefore, the
velocity of a semiclassical carrier constructed from an effective quantum
theory in the presence of external electric field and static disorder reads
\begin{equation}
\mathbf{v}_{l}=\mathbf{\tilde{v}}_{l}^{0}+\mathbf{\tilde{v}}_{l}%
^{sj}+\mathbf{\tilde{v}}_{l}^{B}, \label{velocity}%
\end{equation}
which contributes to the electrical current as $\mathbf{j}=e\sum_{l}%
\mathbf{v}_{l}f_{l}$. The intrinsic anomalous Hall current reads%
\begin{equation}
\mathbf{j}^{in}=e\sum_{l}f_{l}^{0}\mathbf{\tilde{v}}_{l}^{B},
\end{equation}
and the anomalous quantum contribution is
\begin{equation}
\mathbf{j}^{AQ}=\mathbf{j}^{sj}+\mathbf{j}^{ad}+\mathbf{j}^{pair},
\end{equation}
where $\mathbf{j}^{sj}=e\sum_{l}g_{l}^{\left(  -2\right)  }\mathbf{\tilde{v}%
}_{l}^{sj}$,\ $\mathbf{j}^{ad}=e\sum_{l}g_{l}^{a}\mathbf{\tilde{v}}_{l}^{0}$
and $\mathbf{j}^{pair}=e\sum_{l}g_{l}^{pair}\mathbf{\tilde{v}}_{l}^{0}$. The
anomalous quantum mechanism comprises three contributions
\cite{Nagaosa2010,Sinitsyn2007}: (1) a component of the side-jump velocity
transverse to the driving electric field; (2) an anomalous distribution
function \cite{Sinitsyn2005,Sinitsyn2006}; (3) scattering off pairs of
impurities. Both $\mathbf{\tilde{v}}_{l}^{sj}$ and $g_{l}^{a}$ are related to
the coordinate-shift \cite{Sinitsyn2005}, i.e., the so-called side-jump
transverse to the incident wave-vector \cite{Berger1970}.\ Besides, the skew
scattering contribution reads%
\begin{equation}
\mathbf{j}^{sk}=e\sum_{l}g_{l}^{sk}\mathbf{\tilde{v}}_{l}^{0}.
\end{equation}

\section{Application 1: 2D Nozieres-Lewiner effective model}

As the first application, we look into the spin Hall effect in the
widely-studied Nozieres-Lewiner effective model where $\hat{H}_{0}=\frac
{\hbar^{2}\mathbf{\hat{k}}^{2}}{2m}$ in Eq. (\ref{model})
\cite{1973,Sinova2015,Engel2005,Tse2006PRL,Vignale2006PRB,Vignale2009JPCM,Culcer2010}%
. For simplicity we only consider 2D electrons, and the random scalar disorder
is modeled by $\left\langle V\left(  \mathbf{r}\right)  V\left(
\mathbf{r}^{\prime}\right)  \right\rangle =n_{im}V_{0}^{2}\delta\left(
\mathbf{r}-\mathbf{r}^{\prime}\right)  $ with $n_{im}$ the impurity density
and $V_{0}$ the scattering amplitude. The skew scattering is thus neglected in
this section, because it has been thoroughly understood \cite{Nagaosa2017}.
The z-component of spin is conserved and the two spin channels are
independent. In this model $\mathbf{J}_{l}=0$, $i\mathbf{J}_{l^{\prime}}%
^{a}-i\mathbf{J}_{l}^{a}=-\mathbf{\hat{D}}\arg W_{l^{\prime}l}=\frac
{\eta\lambda_{0}^{2}}{4}\mathbf{\hat{z}}\times\left(  \mathbf{k}^{\prime
}-\mathbf{k}\right)  $ where $\left(  \hat{\sigma}_{z}\right)  _{\eta\eta
}=\eta$, and%
\begin{equation}
\delta\mathbf{\tilde{r}}_{l^{\prime}l}=2\left(  i\mathbf{J}_{l^{\prime}}%
^{a}-i\mathbf{J}_{l}^{a}\right)  =-2\mathbf{\hat{D}}\arg W_{l^{\prime}l}.
\label{shift-NL}%
\end{equation}
At the same time $\delta^{sj}\mathbf{v}_{l}=0$, $\delta^{in}\mathbf{v}_{l}=0$
and%
\[
\mathbf{\tilde{v}}_{l}^{sj}=\delta^{sj,1}\mathbf{v}_{l}=\sum_{l^{\prime}%
}\omega_{ll^{\prime}}^{\left(  2\right)  }\delta\mathbf{\tilde{r}}_{l^{\prime
}l},\text{ }\mathbf{\tilde{v}}_{l}^{B}=\delta^{in,1}\mathbf{v}_{l}=\Omega
_{l}^{a}\mathbf{\hat{z}}\times\frac{e}{\hbar}\mathbf{E}.
\]
Simple derivations yield $-\mathbf{j}^{z,in}=\mathbf{j}^{z,sj}=\mathbf{j}%
^{z,ad}$, and $\mathbf{j}^{z,pair}=0$ in the noncrossing approximation, thus
the total $o\left(  n_{im}^{0}\right)  $ contribution is $\mathbf{j}%
^{z}=\mathbf{j}^{z,ad}$.

A phenomenological Boltzmann approach has been employed to analyze this model
\cite{1973,Vignale2006PRB,Vignale2009,Vignale2009JPCM}. In that alternative
Boltzmann approach, there are three basic pictorial arguments: (1) The
contributions from $\delta^{V}\mathbf{\hat{v}}$ and $\delta^{\mathbf{E}%
}\mathbf{\hat{v}}$ to $\left\langle \mathbf{\hat{v}}\right\rangle _{av}$
cancel out (here $\left\langle \mathbf{\hat{v}}\right\rangle _{av}%
=\sum_{\mathbf{k}}f_{\eta\mathbf{k}}\mathbf{v}_{\eta\mathbf{k}}$), for
$\delta^{V}\mathbf{\hat{v}}+\delta^{\mathbf{E}}\mathbf{\hat{v}}=2\frac
{\lambda_{0}^{2}}{4\hbar}\hat{\sigma}_{z}\left(  \mathbf{\nabla}\hat
{V}-e\mathbf{E}\right)  \times\mathbf{\hat{z}}$ and the \textquotedblleft net
force\textquotedblright\ $e\mathbf{E}-\mathbf{\nabla}\hat{V}=\hbar
d\mathbf{\hat{k}/}dt$ acting on electrons equals zero in\textit{ steady-states
}$\left\langle d\mathbf{\hat{k}/}dt\right\rangle _{av}=0$. (2) The $\delta
$-function $\delta\left(  \epsilon_{\mathbf{k}^{\prime}}^{\eta}-\epsilon
_{\mathbf{k}}^{\eta}\right)  $ for the energy-conservation in the
semiclassical scattering rate is replaced by $\delta\left(  \epsilon
_{\mathbf{k}^{\prime}}^{\eta}-\epsilon_{\mathbf{k}}^{\eta}-e\mathbf{E}%
\cdot\Delta\mathbf{r}_{phy}\right)  $ in the presence of the electric field
and SOI. (3) The shift of the physical position $\Delta\mathbf{r}_{phy}$
during a specific scattering process from $l=\left(  \eta,\mathbf{k}\right)  $
state to $l^{\prime}=\left(  \eta,\mathbf{k}^{\prime}\right)  $\ state is
obtained by a time-dependent analysis: $\Delta\mathbf{r}_{phy}=\int
dt\mathbf{v}_{phy}$, and the nontrivial part of $\Delta\mathbf{r}_{phy}$ that
survives after average over many scatterings is $\Delta\mathbf{r}_{phy}%
=2\frac{\lambda_{0}^{2}}{4}\left(  \hat{\sigma}_{z}\right)  _{\eta\eta
}\mathbf{\hat{z}}\times\int dt\frac{d\mathbf{k}}{dt}=2\frac{\lambda_{0}^{2}%
}{4}\eta\mathbf{\hat{z}}\times\left(  \mathbf{k}^{\prime}-\mathbf{k}\right)
$, due to $\mathbf{\hat{v}}=\frac{\hbar\mathbf{\hat{k}}}{m}+2\frac{\lambda
_{0}^{2}}{4}\hat{\sigma}_{z}\frac{d\mathbf{\hat{k}}}{dt}\times\mathbf{\hat{z}%
}$.

Now we point out the correspondence of these three points in the present SB
theory. First, point (1) is equivalent to our $\mathbf{j}^{z,in}%
=-\mathbf{j}^{z,sj}$. As will be detailed in Sec. VI and Appendix D, the
argument in above point (1) is not always valid when $\hat{\sigma}_{z}$ has
$l$-off-diagonal matrix elements. Second, the energy-conservation in the
scattering process in the presence of the electric field and SOI is accounted
for in the present formalism in another way different from above point (2).
The effect of electric-field working $e\mathbf{E}\cdot\Delta\mathbf{r}_{phy}$
during the scattering is compensated by introducing the anomalous distribution
function, and the $\delta$-function $\delta\left(  \epsilon_{\mathbf{k}%
^{\prime}}^{\eta}-\epsilon_{\mathbf{k}}^{\eta}\right)  $ for the
energy-conservation in the semiclassical scattering rate remains unchanged
\cite{Sinitsyn2006,Sinitsyn2008,Nagaosa2010}. Finally, $\Delta\mathbf{r}%
_{phy}$ obtained in above point (3) is equal to our $\delta\mathbf{\tilde{r}%
}_{l^{\prime}l}=2\left(  i\mathbf{J}_{l^{\prime}}^{a}-i\mathbf{J}_{l}%
^{a}\right)  $, with the important factor 2.

Regarding this model, there are some more theoretical debates worthwhile to be
clarified. \begin{figure}[b]
\includegraphics[width=1\columnwidth]{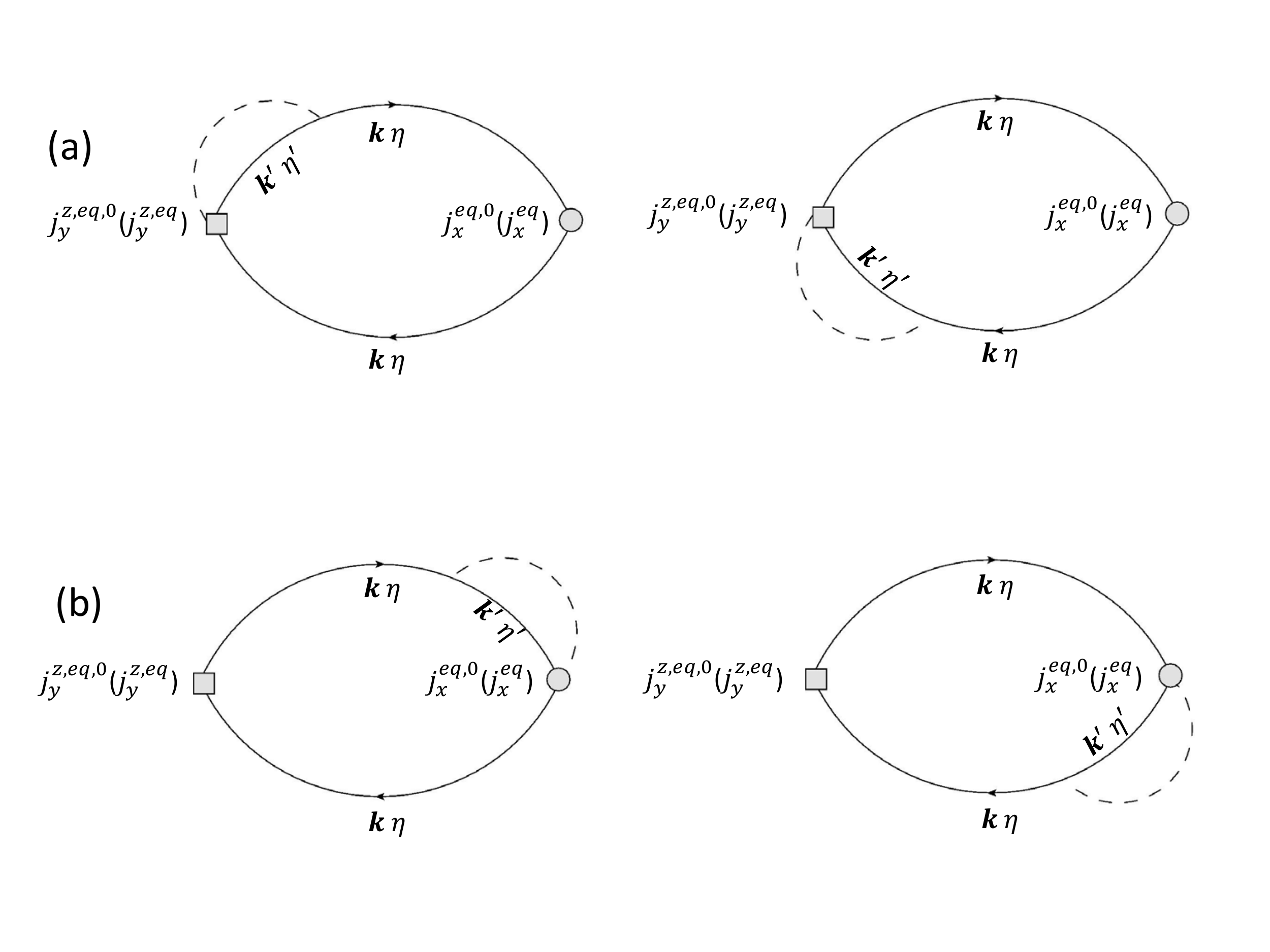}\caption{Side-jump diagrams for
the spin Hall conductivity in the Nozieres-Lewiner effective model. Squares
and circles represent spin-current ($j^{z,eq}$) and charge-current ($j^{eq}$)
vertexes respectively. $j^{z,eq,0}$ and $j^{eq,0}$ are vertexes in the usual
Kubo-Streda formalism (Eq. (\ref{Kubo-usual})) whereas $j^{z,eq}$ and $j^{eq}$
are vertexes in the generalized Kubo-Streda formalism (Eq.
(\ref{Kubo-effective})). (a) and (b) are spin Hall conductivities arising from
the side-jump spin-current and anomalous distribution function respectively
discussed in details in the main text. }%
\end{figure}

First, if one starts from the effective Hamiltonian (2) with $\hat{H}%
_{0}=\frac{\hbar^{2}\mathbf{\hat{k}}^{2}}{2m}$ but neglect the change of the
physical position operator, then only one half of $\delta\mathbf{\tilde{r}%
}_{l^{\prime}l}$ and thus of $\mathbf{j}^{z,ad}$ can be produced in the SB
theory. This indicates the fact that the effective Hamiltonian itself is not
always enough for consistent predictions with the complete effective quantum
theory \cite{Chang2008,Niu2010}.

Second, in an early influential work, Lyo and Holstein \cite{Lyo1972} claimed
that Berger's wavepacket analysis on the side-jump \cite{Berger1970} is
consistent with the original Luttinger theory \cite{Luttinger1958}, if in the
latter the scalar disorder potential $V\left(  \mathbf{r}\right)  $ is
replaced by $W$. Taking into account the scattering correction to the
plane-wave state ($\hat{H}_{0}=\frac{\hbar^{2}\mathbf{\hat{k}}^{2}}{2m}$) in
the lowest Born order, Lyo-Holstein in fact proved%
\[
\delta_{LH}^{sj}\mathbf{v}_{l}\equiv2\operatorname{Re}\sum_{l^{\prime}\neq
l}\frac{\left\langle \left(  \frac{1}{i\hbar}\left[  \mathbf{\hat{r}},\hat
{W}\right]  \right)  _{ll^{\prime}}W_{l^{\prime}l}\right\rangle }%
{d_{ll^{\prime}}^{+}}=\sum_{l^{\prime}}\omega_{ll^{\prime}}^{\left(  2\right)
}\delta\mathbf{r}_{l^{\prime}l},
\]
where $\delta\mathbf{r}_{l^{\prime}l}=-\mathbf{\hat{D}}\arg W_{l^{\prime}l}$.
And $\delta_{LH}^{sj}\mathbf{v}_{l}\tau_{l}^{tr}=\frac{\eta\lambda_{0}^{2}}%
{4}\mathbf{k\times\hat{z}}$ gives the sideways shift of Berger
\cite{Berger1970}, with $\tau_{l}^{tr}$ the transport time. According to Eq.
(\ref{shift-NL}), the Lyo-Holstein result is nothing else but%
\[
\delta_{LH}^{sj}\mathbf{v}_{l}=\frac{1}{2}\delta^{sj,1}\mathbf{v}_{l}.
\]
The other half of $\delta^{sj,1}\mathbf{v}_{l}$ did not appear in the
Lyo-Holstein theory, because the change of the physical position operator was
not considered there. The argument by Lyo-Holstein under the weak
disorder-potential Born approximation has been generalized directly to the
level of full T-matrix \cite{Levy}. But that generalization has not been
confirmed by microscopic quantum transport theories.

Third, the usual Kubo-Streda theory applied to the present model
\cite{Bruno2001,Dugaev2001,Tse2006PRL} in fact starts only from the
Hamiltonian $\hat{H}=\frac{\hbar^{2}\mathbf{\hat{k}}^{2}}{2m}+\hat
{W}-e\mathbf{E\cdot\hat{r}}$. Figure 1 with the current vertexes
$j_{y}^{z,eq,0}$ and $j_{x}^{eq,0}$ shows the four well-known side-jump
diagrams in the diagrammatic approach for the present model. Hereafter the
external electric field is applied in x direction. In the language of the SB
theory, figure 1 (a) and 1 (b) (with the current vertexes $j_{y}^{z,eq,0}$ and
$j_{x}^{eq,0}$) represent the spin Hall conductivities $\sigma_{yx}%
^{z,0,sj}=e\sum_{l}\frac{\hbar}{2}\eta\left(  \delta_{LH}^{sj}\mathbf{v}%
_{l}\right)  _{y}\left(  -\frac{\partial f^{0}}{\partial\epsilon_{l}}\right)
\tau_{l}^{tr}\left(  \mathbf{v}_{l}^{0}\right)  _{x}$ and $\sigma
_{yx}^{z,0,ad}=e\sum_{l}\frac{\hbar}{2}\eta\left(  \mathbf{v}_{l}^{0}\right)
_{y}\frac{\partial f^{0}}{\partial\epsilon_{l}}\tau_{l}^{tr}\left(
\delta_{LH}^{sj}\mathbf{v}_{l}\right)  _{x}$, respectively. Thus $\sigma
_{yx}^{z,0,sj}=\sigma_{yx}^{z,0,ad}=\frac{1}{2}\sigma_{yx}^{z,ad}=\frac{1}%
{2}\sigma_{yx}^{z,sj}$, and the total $o\left(  n_{im}^{0}\right)  $
contribution $\sigma_{yx}^{z,0,sj}+\sigma_{yx}^{z,0,ad}=\sigma_{yx}^{z,ad}$
coincides with our SB result. However, the physical interpretation given by
this diagrammatic approach is very different from that obtained in the quantum
kinetic \cite{Culcer2010} or the SB theories. Being important in the quantum
kinetic theory, the SOI with the driving electric field is not incorporated
into the derivation of the usual Kubo-Streda formula
\cite{Streda2010,Bruno2001}. Thus for spin-orbit-induced transport in
effective quantum theories, the SB and quantum kinetic theories do not always
produce the same result as the usual Kubo-Streda fomula. In the next section
we will provide such an example.

Moreover, in Appendix C we give the slightly generalized Kubo-Streda formula
in effective quantum theories, taking into account the change of the physical
position operator. That formula yields the same physical picture as the SB
theory. The expressions for the current vertex in the corresponding diagrams
(Eq. (\ref{Kubo-effective})) are different from those in the usual diagrams
(Eq. (\ref{Kubo-usual})), see Appendix C. Besides, the generalized Kubo-Streda
formula contains an additional \textquotedblleft Fermi sea\textquotedblright%
\ term (Eq. (\ref{additional})) which results from $\delta^{in,1}%
\mathbf{j}_{l}^{z}$.

\section{Application 2: Rashba 2D effective model}

In the present case $\hat{H}_{0}=\frac{\hbar^{2}\mathbf{\hat{k}}^{2}}%
{2m}+\alpha_{R}\mathbf{\hat{\sigma}}\cdot\left(  \mathbf{\hat{k}}%
\times\mathbf{\hat{z}}\right)  $ in the effective Hamiltonian (\ref{model}),
$\alpha_{R}$ is the Rashba SOI parameter. The Rashba spinor in the internal
space reads $|u_{\mathbf{k}}^{\eta}\rangle=\frac{1}{\sqrt{2}}\left(
1,-i\eta\exp\left(  i\phi\right)  \right)  ^{T}$. Although the Rashba SOI can
be understood as arising also from $\mathbf{\hat{r}}^{a}$, we here take a more
general viewpoint beyond the context of direct-gap semiconductors, regarding
this model as an example where the effective Hamiltonian has its own internal
structure. In this model there is no unique definition of the spin current
operator \cite{Shi2006}. For the convenience of making comparison with most
previous researches on this model
\cite{Tse2006PRB,Vignale2008,Raimondi2012,Culcer2013,Shen2014}, the
conventional definition of spin current operator is adopted in the present
paper. The results on the conserved spin current proposed in Ref.
\cite{Shi2006} will be presented elsewhere.

The spin current polarized in z direction is $\mathbf{\hat{\jmath}}^{z}%
=\frac{1}{2}\left\{  \frac{\hbar}{2}\hat{\sigma}_{z},\mathbf{\hat{v}}\right\}
=\mathbf{\hat{\jmath}}^{z,phy}+\delta^{V}\mathbf{\hat{\jmath}}^{z}%
+\delta^{\mathbf{E}}\mathbf{\hat{\jmath}}^{z}$, where $\delta^{V}%
\mathbf{\hat{\jmath}}^{z}+\delta^{\mathbf{E}}\mathbf{\hat{\jmath}}^{z}%
=\frac{\lambda_{0}^{2}}{4}\mathbf{\hat{z}}\times\left(  e\mathbf{E}%
-\mathbf{\nabla}\hat{V}\right)  $. In this model $\mathbf{\hat{\jmath}%
}^{z,phy}=\frac{\hbar}{2}\frac{\hbar\mathbf{k}}{m}\hat{\sigma}_{z}$, then
$\mathbf{\tilde{j}}_{l}^{z,0}=0$, thus $\mathbf{j}^{z,sk}=\mathbf{j}%
^{z,pair}=\mathbf{j}^{z,ad}=0$ and%
\[
\mathbf{j}^{z}=\sum_{l}f_{l}^{0}\left(  \delta^{in}\mathbf{j}_{l}^{z}%
+\delta^{in,1}\mathbf{j}_{l}^{z}\right)  +\sum_{l}g_{l}^{\left(  2\right)
}\left(  \delta^{sj}\mathbf{j}_{l}^{z}+\delta^{sj,1}\mathbf{j}_{l}^{z}\right)
.
\]
As will be explained in Sec. VI, we have $\sum_{l}f_{l}^{0}\delta
^{in,1}\mathbf{j}_{l}^{z}=-\sum_{l}g_{l}^{\left(  2\right)  }\delta
^{sj,1}\mathbf{j}_{l}^{z}$, then
\begin{equation}
\mathbf{j}^{z}=\sum_{l}f_{l}^{0}\delta^{in}\mathbf{j}_{l}^{z}+\sum_{l}%
g_{l}^{\left(  2\right)  }\delta^{sj}\mathbf{j}_{l}^{z},
\end{equation}
where $\delta^{in}\mathbf{j}_{l}^{z}=-2e\mathbf{E}\cdot\operatorname{Re}%
\frac{i\mathbf{\tilde{J}}_{\eta\mathbf{k},-\eta\mathbf{k}}\langle
u_{\mathbf{k}}^{-\eta}|\mathbf{\hat{\jmath}}^{z,0}|u_{\mathbf{k}}^{\eta
}\rangle}{\epsilon_{\mathbf{k}}^{\eta}-\epsilon_{\mathbf{k}}^{-\eta}}$ and
$\delta^{sj}\mathbf{j}_{l}^{z}$ is given by Eq. (\ref{sj}).

We only consider the case of both subbands partially occupied. The
corresponding wave number in $\eta$ band is given as $k_{\eta}\left(
\epsilon\right)  =-\eta k_{R}+\alpha_{R}^{-1}\sqrt{\epsilon_{R}^{2}%
+2\epsilon_{R}\epsilon}$. Here $k_{R}=m\frac{\alpha_{R}}{\hbar^{2}}%
=k_{-}\left(  \epsilon\right)  -k_{+}\left(  \epsilon\right)  $ measures the
momentum splitting of two Rashba bands. We assume $\left(  k_{R}/k_{F}\right)
^{2}\gg\left(  \lambda_{0}k_{F}\right)  ^{2}$, i.e., $\left(  \lambda_{0}%
k_{R}\right)  ^{2}\gg\left(  \lambda_{0}k_{F}\right)  ^{4}$, which is the case
of strong Rashba SOI. Here $k_{F}\equiv k_{\eta}\left(  \epsilon_{F}\right)
+\eta k_{R}$. The density of state of $\eta$ band takes the form $D_{\eta
}\left(  \epsilon\right)  =D_{0}\frac{k_{\eta}\left(  \epsilon\right)
}{k_{\eta}\left(  \epsilon\right)  +\eta k_{R}}$, with $D_{0}=\frac{m}%
{2\pi\hbar^{2}}$.

Equation (\ref{sj})\ yields $\delta^{sj}\mathbf{j}_{l}^{z}=\frac{\eta}%
{4k_{R}\tau_{0}}\frac{\hbar}{2}\mathbf{\hat{e}}_{\mathbf{k}}\times
\mathbf{\hat{z}}$ with $\tau_{0}^{-1}=\frac{2\pi}{\hbar}n_{im}V_{0}^{2}D_{0}$,
$\mathbf{\hat{e}}_{\mathbf{k}}=\left(  \cos\phi,\sin\phi\right)  $. The
longitudinal distribution function is $g_{l}^{\left(  -2\right)  }%
=eE_{x}\left(  -\partial_{\epsilon_{l}}f^{0}\right)  \left(  \mathbf{v}%
_{l}^{0}\right)  _{x}\tau_{l}^{tr}$, where
\begin{equation}
\frac{\tau_{\eta}^{tr}\left(  \epsilon\right)  }{\tau_{0}}=\frac{D_{\eta}%
}{D_{0}}-\eta\frac{\lambda_{0}^{2}}{4}4k_{R}k_{-\eta}+2\frac{\lambda_{0}^{2}%
}{4}\left(  2k_{R}\right)  ^{2}\frac{D_{\eta}}{D_{0}}. \label{transport-time}%
\end{equation}
Thus%
\begin{equation}
\frac{\sum_{l}f_{l}^{0}\left(  \delta^{in}\mathbf{j}_{l}^{z}\right)  _{y}%
}{E_{x}}=\frac{-e}{8\pi}-\frac{en_{e}\lambda_{0}^{2}}{4}\left(  \frac{1}%
{2}-\frac{\frac{1}{6}\epsilon_{R}}{\epsilon_{F}+\epsilon_{R}}\right)  ,
\label{Culcer-2}%
\end{equation}%
\begin{equation}
\frac{\sum_{l}g_{l}^{\left(  -2\right)  }\left(  \delta^{sj}\mathbf{j}_{l}%
^{z}\right)  _{y}}{E_{x}}=\frac{e}{8\pi}+\frac{en_{e}\lambda_{0}^{2}}%
{4}\left(  \frac{1}{2}+\frac{\frac{1}{2}\epsilon_{R}}{\epsilon_{F}%
+\epsilon_{R}}\right)  , \label{Culcer-1}%
\end{equation}
and then%
\begin{equation}
j_{y}^{z}=\frac{en_{e}\lambda_{0}^{2}}{4}\frac{\frac{2}{3}\epsilon_{R}%
}{\epsilon_{F}+\epsilon_{R}}E_{x}=\left(  \lambda_{0}k_{R}\right)  ^{2}%
\frac{e}{6\pi}E_{x}. \label{SHC}%
\end{equation}
This nonzero spin Hall conductivity of order $\left(  \lambda_{0}k_{R}\right)
^{2}$ did not appear in previous theoretical researches
\cite{Vignale2008,Raimondi2012,Culcer2013} assuming weak Rashba SOI. Strong
Rashba SOI energy compared to the Fermi energy is possible, e.g., in
heterostructures of non-centrosymmetric semiconductor BiTeX (X=Cl, Br and I)
family \cite{Wu2014}. Because we do not assume $k_{R}\ll k_{F}$, in the case
of strong Rashba SOI the magnitude of the above spin Hall conductivity can be
comparable to the side-jump spin Hall conductivity in the 2D Nozieres-Lewiner model.

The SB formalism provides alternative explanations to the results in a recent
quantum kinetic theory \cite{Culcer2013}. The second term on the rhs of Eq.
(\ref{Culcer-2}) arises from the SOI with the external driving electric field,
and corresponds to the contribution called \textquotedblleft anomalous spin
precession from electric field\textquotedblright\ in Ref. \cite{Culcer2013}.
Whereas the second term on the rhs of Eq. (\ref{Culcer-1}) arises from the
spin-orbit-scattering-induced correction to the transport time in Eq.
(\ref{transport-time}), and corresponds to the contribution called
\textquotedblleft anomalous spin precession from impurities\textquotedblright%
\ in Ref. \cite{Culcer2013}. The $o\left(  \left(  \lambda_{0}k_{R}\right)
^{2}\right)  $ contribution is neglected in Ref. \cite{Culcer2013}, where the
weak Rashba SOI was assumed.

\begin{figure}[b]
\includegraphics[width=1\columnwidth]{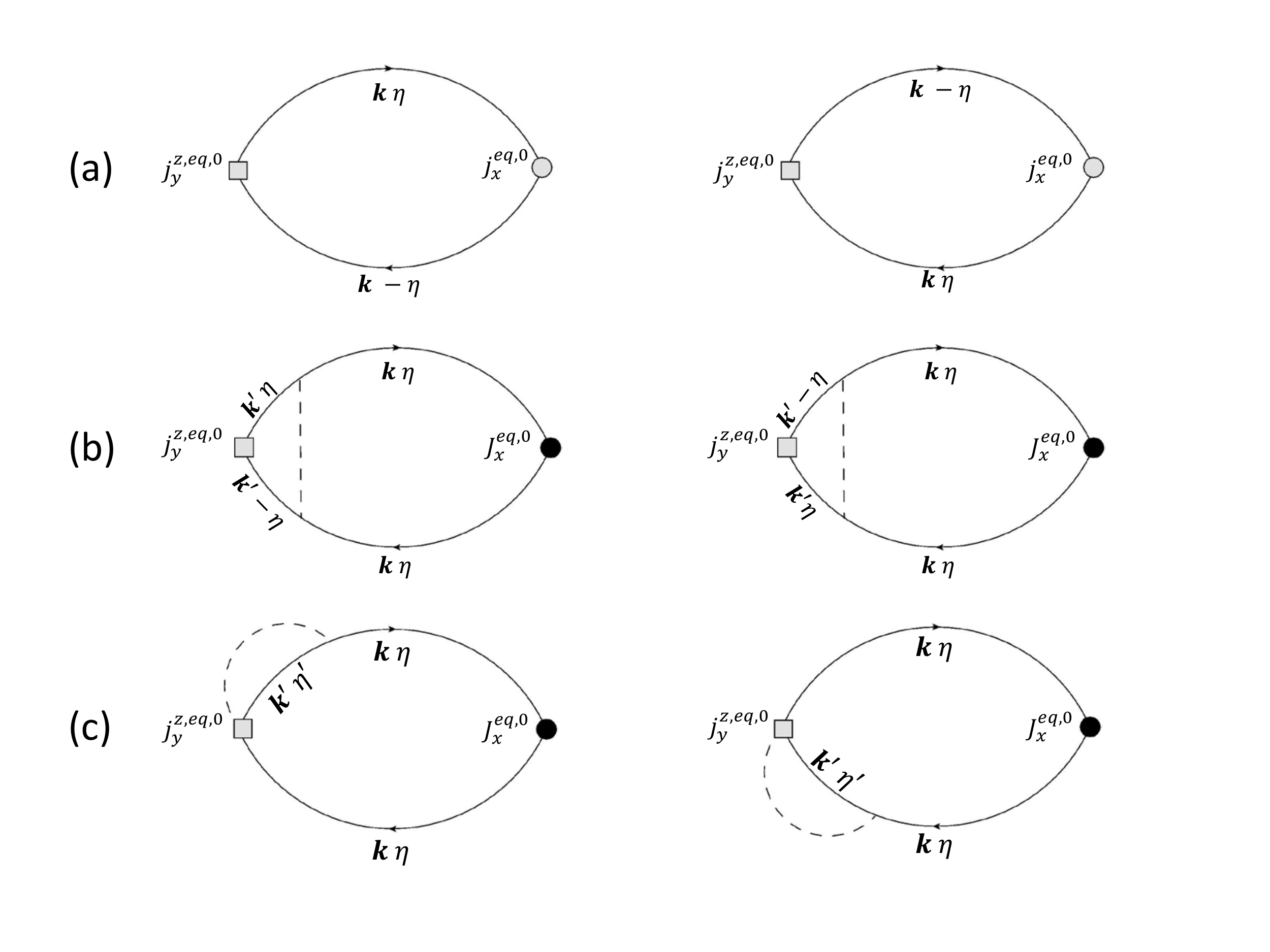} \caption{Diagrams for the
non-zero contributions to the spin Hall conductivity in the usual Kubo-Streda
formalism. Squares represent spin-current ($j^{z,eq,0}$) vertexes. Circles and
filled circles represent bare charge-current ($j^{eq,0}$) and renormalized
charge-current ($J^{eq,0}$) vertexes respectively. (a) describes intrinsic
contributions; (b) describes side-jump contributions from interband coherence
response due to band structure SOI; (c) describes side-jump contributions from
disorder-induced corrections to the charge-current operator. }%
\label{Fig:2}%
\end{figure}Now we consider the application of the usual Kubo-Streda formula,
which in fact treats the Hamiltonian $\hat{H}=\hat{H}_{0}+\hat{W}%
-e\mathbf{E\cdot\hat{r}}$, as in Ref. \cite{Tse2006PRB}. In the weak
disorder-potential regime the diagram calculation can be done in the
band-eigenstate representation, where the correspondence to the SB formalism
is apparent \cite{Sinitsyn2007,Nagaosa2010,Xiao2017SOT-SBE}. In the language
of the SB theory, the diagrams in Fig. 2 yield
\begin{equation}
\sigma_{yx}^{z,0}=\sum_{l}f_{l}^{0}\frac{\left(  \delta_{0}^{in}\mathbf{j}%
_{l}^{z}\right)  _{y}}{E_{x}}+\sum_{l}\left(  \delta_{0}^{sj}\mathbf{j}%
_{l}^{z}+\delta_{0}^{sj,1}\mathbf{j}_{l}^{z}\right)  _{y}\frac{g_{l}^{\left(
2\right)  }}{E_{x}}, \label{diagram}%
\end{equation}
where $\delta_{0}^{in}\mathbf{j}_{l}^{z}=-2e\mathbf{E}\cdot\operatorname{Re}%
\frac{i\mathbf{J}_{\eta\mathbf{k},-\eta\mathbf{k}}\langle u_{\mathbf{k}%
}^{-\eta}|\mathbf{\hat{\jmath}}^{z,0}|u_{\mathbf{k}}^{\eta}\rangle}%
{\epsilon_{\mathbf{k}}^{\eta}-\epsilon_{\mathbf{k}}^{-\eta}}$ and
\[
\delta_{0}^{sj}\mathbf{j}_{l}^{z}=\sum_{l^{\prime},l^{\prime\prime}\neq
l^{\prime}}\frac{\left\langle W_{ll^{\prime}}W_{l^{\prime\prime}%
l}\right\rangle \mathbf{j}_{l^{\prime}l^{\prime\prime}}^{z,0}}{d_{ll^{\prime}%
}^{-}d_{ll^{\prime\prime}}^{+}}+2\operatorname{Re}\sum_{l^{\prime}\neq
l,l^{\prime\prime}}\frac{\left\langle W_{l^{\prime}l^{\prime\prime}%
}W_{l^{\prime\prime}l}\right\rangle \mathbf{j}_{ll^{\prime}}^{z,0}%
}{d_{ll^{\prime}}^{+}d_{ll^{\prime\prime}}^{+}},
\]%
\[
\delta_{0}^{sj,1}\mathbf{j}_{l}^{z}=2\operatorname{Re}\sum_{l^{\prime}\neq
l}\frac{\left\langle \langle l|\frac{1}{2}\left\{  \frac{\hbar}{2}\hat{\sigma
}_{z},\frac{1}{i\hbar}\left[  \mathbf{\hat{r}},\hat{W}\right]  \right\}
\mathbf{|}l^{\prime}\rangle W_{l^{\prime}l}\right\rangle }{d_{ll^{\prime}}%
^{+}}.
\]
The first term on the rhs of Eq. (\ref{diagram}) corresponds to Fig. 2 (a)
($\sigma_{yx}^{z,II,0}=0$ in the considered model, see Appendix C), whereas
the contributions from $\delta_{0}^{sj}\mathbf{j}_{l}^{z}$ and $\delta
_{0}^{sj,1}\mathbf{j}_{l}^{z}$ in the second term on the rhs of Eq.
(\ref{diagram}) correspond to Fig. 2 (b) and 2 (c), respectively. One can find
that $\delta_{0}^{in}\mathbf{j}_{l}^{z}\neq\delta^{in}\mathbf{j}_{l}^{z}$,
$\delta_{0}^{sj}\mathbf{j}_{l}^{z}=\delta^{sj}\mathbf{j}_{l}^{z}$, $\delta
_{0}^{sj,1}\mathbf{j}_{l}^{z}=\frac{1}{2}\delta^{sj,1}\mathbf{j}_{l}^{z}$ up
to the first order of $\lambda_{0}^{2}$ for the present model. Regarding the
spin Hall conductivity $j_{y}^{z}/E_{x}$ we have $\sum_{l}f_{l}^{0}%
\frac{\left(  \delta_{0}^{in}\mathbf{j}_{l}^{z}\right)  _{y}}{E_{x}}=\frac
{-e}{8\pi}$, $\sum_{l}\left(  \delta_{0}^{sj,1}\mathbf{j}_{l}^{z}\right)
_{y}\frac{g_{l}^{\left(  2\right)  }}{E_{x}}=-\frac{1}{2}\frac{en_{e}%
\lambda_{0}^{2}}{4}$, and $\sum_{l}\left(  \delta_{0}^{sj}\mathbf{j}_{l}%
^{z}\right)  _{y}\frac{g_{l}^{\left(  2\right)  }}{E_{x}}$ is the same as Eq.
(\ref{Culcer-1}). In the calculation we used $\left(  \delta_{0}%
^{sj,1}\mathbf{j}_{l}^{z}\right)  _{y}=-\frac{\hbar}{2}\frac{\lambda_{0}^{2}%
}{4\tau_{0}}\left(  k_{\eta}+\eta k_{R}\right)  \cos\phi$. Then $\sigma
_{yx}^{z,0}=\left(  \lambda_{0}k_{R}\right)  ^{2}\frac{e}{8\pi}$. This result
differs from Eq. (\ref{SHC}) obtained for the complete effective model.

\section{Discussion and Summary}

We discuss the validity of the widely accepted idea that the effects from
corrections to the current operator due to the driving electric field and the
gradient of disorder potential always cancel out
\cite{Vignale2006PRB,Vignale2009JPCM,Culcer2010,Culcer2013}. This idea
indicates $\sum_{l}g_{l}^{\left(  -2\right)  }\delta^{sj,1}\mathbf{v}%
_{l}=-\sum_{l}f_{l}^{0}\delta^{in,1}\mathbf{v}_{l}$ and $\sum_{l}%
g_{l}^{\left(  -2\right)  }\delta^{sj,1}\mathbf{j}_{l}^{z}=-\sum_{l}f_{l}%
^{0}\delta^{in,1}\mathbf{j}_{l}^{z}$. A pictorial argument of this idea is
based on the correspondence principle (Ehrenfest's theorem)
\cite{Vignale2006PRB,Vignale2009JPCM,Culcer2010}, as mentioned in Sec. IV. For
the Nozieres-Lewiner effective model ($\hat{H}_{0}=\frac{\hbar^{2}%
\mathbf{\hat{k}}^{2}}{2m}$), this idea has been confirmed by a quantum kinetic
theory \cite{Culcer2010}. However, notice that $\hat{\sigma}_{z}$ is also
involved in $\delta^{V}\mathbf{\hat{v}}+\delta^{\mathbf{E}}\mathbf{\hat{v}%
}=2\frac{\lambda_{0}^{2}}{4\hbar}\hat{\sigma}_{z}\left(  \mathbf{\nabla}%
\hat{V}-e\mathbf{E}\right)  \times\mathbf{\hat{z}}$ (for 2D electrons). When
the off-diagonal matrix elements of $\hat{\sigma}_{z}$ in the band-eigenstate
basis is not zero, the Ehrenfest's theorem cannot be employed directly. Thus
$\sum_{l}g_{l}^{\left(  -2\right)  }\delta^{sj,1}\mathbf{v}_{l}=-\sum_{l}%
f_{l}^{0}\delta^{in,1}\mathbf{v}_{l}$ is not generally valid even for 2D
electrons. In Appendix D we confirm this observation by the SB formalism.

On the other hand, considering the conventional spin current polarized in z
direction for 2D electrons, $\delta^{V}\mathbf{\hat{\jmath}}^{z}%
+\delta^{\mathbf{E}}\mathbf{\hat{\jmath}}^{z}=\frac{\lambda_{0}^{2}}%
{4}\mathbf{\hat{z}}\times\left(  e\mathbf{E}-\mathbf{\nabla}\hat{V}\right)  $
is free from the above problem. Thus the Ehrenfest's theorem can be applied,
validating the pictorial argument \cite{Vignale2006PRB} for 2D electrons even
with band-structure SOI. Specifically, one has $\sum_{l}f_{l}^{0}\left(
\delta^{in,1}\mathbf{j}_{l}^{z}\right)  _{y}=E_{x}\frac{e\lambda_{0}^{2}}%
{4}\sum_{l}f_{l}^{0}=E_{x}\frac{en_{e}\lambda_{0}^{2}}{4}$ and $\sum_{l}%
g_{l}^{\left(  -2\right)  }\left(  \delta^{sj,1}\mathbf{j}_{l}^{z}\right)
_{y}=2\frac{\hbar}{2}\sum_{ll^{\prime}}g_{l}^{\left(  -2\right)  }%
\omega_{l^{\prime}l}^{\left(  2\right)  }\frac{\lambda_{0}^{2}}{4}\left(
k_{x}^{\prime}-k_{x}\right)  =\frac{\lambda_{0}^{2}}{4}eE_{x}\sum_{l}%
\partial_{k_{x}}f^{0}k_{x}=-\frac{en_{e}\lambda_{0}^{2}}{4}E_{x}$, where we
use the same manipulations as in Eq. (\ref{Vignale-2}).

We note that the special form $\mathbf{\hat{r}}^{a}=\frac{\lambda_{0}^{2}}%
{4}\hat{\sigma}_{z}\mathbf{\hat{z}\times\hat{k}}$ which is linear in the
momentum for 2D electrons plays the vital role in the above derivation. But
for the conventional spin current in 2D hole systems where $\mathbf{\hat{r}%
}^{a}$ is cubic in the momentum, we cannot get the relation $\sum_{l}%
g_{l}^{\left(  -2\right)  }\delta^{sj,1}\mathbf{j}_{l}^{z}=-\sum_{l}f_{l}%
^{0}\delta^{in,1}\mathbf{j}_{l}^{z}$ if no further assumption about the model
Hamiltonian is made. Detailed discussions are presented in Appendix D.

In summary, we constructed the SB transport framework directly in the level of
the effective quantum theory, confirmed by a generalized Kohn-Luttinger
density matrix transport theory also in this level. It was shown that the
spin/anomalous Hall effect studied in this level can still be parsed into the
same categories as in the level of the full Hamiltonian, in the regime where
the SB theory works. We discussed the link and difference between the present
SB theory and various previous theories. To help clarify this issue, we also
derived the slightly generalized Kubo-Streda formula in the level of the
effective quantum theory. This formula leads to the same physical picture as
the present SB theory. In a Rashba 2D effective model, a nonzero spin Hall
effect important in the case of strong Rashba SOI but neglected in previous
theories has been found.

\begin{acknowledgments}
We acknowledge insightful discussions with Q. Niu. C. X. thanks P. Streda and D. Culcer for useful discussions. C. X. and B. X. are supported by DOE (DE-FG03-02ER45958, Division of Materials Science and Engineering), NSF (EFMA-1641101) and Welch Foundation (F-1255).
F. X. is supported by the Department of Energy, Office of Basic Energy Sciences under Contract No. DE-FG02-ER45958 and by the Welch foundation under Grant No. TBF1473.
The formulation in Sec. III is supported by the DOE grant.
\end{acknowledgments}

\appendix

\section{Confirmation of the SB formalism: Kohn-Luttinger density-matrix
approach}

The original Kohn-Luttinger approach is designed in the level of the full
Hamiltonian. Here we formulate a generalized version of it in the level of the
effective quantum theory.

The single-carrier Hamiltonian reads $\hat{H}_{T}=\hat{H}_{0}+\hat{W}+\hat
{H}_{F}$, where the field term $\hat{H}_{F}=\hat{H}_{1}e^{st}$ arises from the
electric field turned on adiabatically from the remote past $t=-\infty$, with
$\hat{H}_{1}=-e\mathbf{E\cdot\hat{r}}^{phy}$. The expectation value of a
single-carrier operator $\hat{A}$ is $\left\langle A\right\rangle =tr\left[
\hat{A}\hat{\rho}_{T}\right]  $, where $tr$ denotes the trace operation in the
single-carrier Hilbert space, and the single-carrier density matrix $\hat
{\rho}_{T}$ is determined by the quantum Liouville equation $i\hbar
\frac{\partial}{\partial t}\hat{\rho}_{T}=\left[  \hat{H}_{T},\hat{\rho}%
_{T}\right]  $. In the linear response regime $\hat{\rho}_{T}=\hat{\rho}%
+\hat{f}e^{st}$, where $\hat{\rho}$ is the equilibrium density matrix,
$\hat{f}$ is linear in the electric field and independent of time. In the
band-eigenstate representation of $\hat{H}_{0}$:
\begin{equation}
d_{ll^{\prime}}^{-}f_{ll^{\prime}}=\left[  \hat{f},\hat{H}^{\prime}\right]
_{ll^{\prime}}+\left[  \hat{\rho},\hat{H}_{1}\right]  _{ll^{\prime}}.
\label{KL}%
\end{equation}
Here $\left[  \hat{A},\hat{B}\right]  _{ll^{\prime}}\equiv\sum_{l^{\prime
\prime}}\left(  A_{ll^{\prime\prime}}B_{l^{\prime\prime}l^{\prime}%
}-B_{ll^{\prime\prime}}A_{l^{\prime\prime}l^{\prime}}\right)  $. Defining
$C_{ll^{\prime}}\equiv\left[  \hat{\rho},\hat{H}_{1}\right]  _{ll^{\prime}}$,
we have $C_{ll^{\prime}}=ie\mathbf{E}\cdot\left[  \mathbf{\hat{D}}%
\rho_{ll^{\prime}}+\left[  \mathbf{\tilde{J}},\rho\right]  _{ll^{\prime}%
}\right]  $ for $l^{\prime}\neq l$ and $C_{l}=ie\mathbf{E}\cdot\left[
\partial_{\mathbf{k}}\rho_{l}+\left[  \mathbf{\tilde{J}},\rho\right]
_{ll}\right]  $. Hereafter $\rho_{ll}\equiv\rho_{l}$, $f_{ll}\equiv f_{l}$ and
$C_{ll}\equiv C_{l}$. In the effective quantum theory, in the presence of the
external electric field $\hat{A}$ may be different from its equilibrium form
$\hat{A}^{eq}$, so $\hat{A}=\hat{A}^{eq}+\delta^{\mathbf{E}}\hat{A}$. The
linear response in the weak disorder-potential regime is thus
\begin{align}
\delta A  &  =tr\left\langle \hat{f}\hat{A}^{eq}\right\rangle +tr\left\langle
\hat{\rho}\delta^{\mathbf{E}}\hat{A}\right\rangle \label{linear-response}\\
&  =\sum_{ll^{\prime}}\left\langle f_{ll^{\prime}}A_{l^{\prime}l}%
^{eq}\right\rangle +\sum_{ll^{\prime}}\left\langle \rho_{ll^{\prime}%
}\right\rangle \left(  \delta^{\mathbf{E}}A\right)  _{l^{\prime}l}\nonumber\\
&  =\sum_{l}\left\langle f_{l}\right\rangle A_{l}^{phy}+\sum_{ll^{\prime}%
}^{\prime}\left\langle f_{ll^{\prime}}A_{l^{\prime}l}^{eq}\right\rangle
+\sum_{l}\rho_{l}^{0}\left(  \delta^{\mathbf{E}}\hat{A}\right)  _{ll}\nonumber
\end{align}
Hereafter the notation $\sum^{\prime}$ means that all the index equalities
should be avoided in the summation.

In the weak disorder-potential regime an iterative solution to Eq. (\ref{KL})
in ascending powers of the disorder potential is possible:
\begin{align*}
f_{l}  &  =f_{l}^{\left(  -2\right)  }+f_{l}^{\left(  -1\right)  }%
+f_{l}^{\left(  0\right)  }+...,\\
f_{ll^{\prime}}  &  =f_{ll^{\prime}}^{\left(  -1\right)  }+f_{ll^{\prime}%
}^{\left(  0\right)  }+f_{ll^{\prime}}^{\left(  1\right)  }...\text{\ }\left(
l\neq l^{\prime}\right)  ,\\
C_{ll^{\prime}}  &  =C_{ll^{\prime}}^{\left(  0\right)  }+C_{ll^{\prime}%
}^{\left(  1\right)  }+C_{ll^{\prime}}^{\left(  2\right)  }+...
\end{align*}
Then one gets an equation for $f_{l}$, as well as expressions for
$f_{ll^{\prime}}\left(  l^{\prime}\neq l\right)  $ in terms of $f_{l}$
\cite{Luttinger1958}:%
\[
f_{ll^{\prime}}^{\left(  -1\right)  }=\frac{f_{l}^{\left(  -2\right)
}-f_{l^{\prime}}^{\left(  -2\right)  }}{d_{ll^{\prime}}^{-}}W_{ll^{\prime}},
\]%
\begin{align*}
f_{ll^{\prime}}^{\left(  0\right)  }  &  =\sum_{l^{\prime\prime}\neq
l,l^{\prime}}\frac{W_{ll^{\prime\prime}}W_{l^{\prime\prime}l^{\prime}}%
}{d_{ll^{\prime}}^{-}}\left[  \frac{f_{l}^{\left(  -2\right)  }-f_{l^{\prime
\prime}}^{\left(  -2\right)  }}{d_{ll^{\prime\prime}}^{-}}-\frac
{f_{l^{\prime\prime}}^{\left(  -2\right)  }-f_{l^{\prime}}^{\left(  -2\right)
}}{d_{l^{\prime\prime}l^{\prime}}^{-}}\right] \\
&  +\frac{f_{l}^{\left(  -1\right)  }-f_{l^{\prime}}^{\left(  -1\right)  }%
}{d_{ll^{\prime}}^{-}}W_{ll^{\prime}}+\frac{C_{ll^{\prime}}^{\left(  0\right)
}}{d_{ll^{\prime}}^{-}}.
\end{align*}
After disorder average one obtains a transport equation for $\left\langle
f_{l}\right\rangle $.\ Comparing Eq. (A3) in Ref. \cite{Xiao2018scaling} to
the present Eq. (\ref{KL}), one can see that they take the same form, except
that $\hat{V}\rightarrow\hat{W}$ and $\mathbf{J}_{ll^{\prime}}\rightarrow
\mathbf{\tilde{J}}_{ll^{\prime}}$. Thus the derivation of the transport
equation for $\left\langle f_{l}\right\rangle $ remains unchanged. Assuming
isotropic systems $\sum_{l^{\prime}}\omega_{l^{\prime}l}^{a}=0$, one has%
\begin{align*}
0  &  =\frac{1}{i\hbar}C_{l}^{\left(  0\right)  }+\sum_{l^{\prime}}%
\omega_{ll^{\prime}}^{\left(  2\right)  }\left\langle f_{l}^{\left(
-2\right)  }-f_{l^{\prime}}^{\left(  -2\right)  }\right\rangle \\
&  +\sum_{l^{\prime}}\omega_{ll^{\prime}}^{\left(  2\right)  }\left\langle
f_{l}^{\left(  -1\right)  }-f_{l^{\prime}}^{\left(  -1\right)  }\right\rangle
+\sum_{l^{\prime}}\omega_{ll^{\prime}}^{\left(  3\right)  }\left\langle
f_{l}^{\left(  -2\right)  }-f_{l^{\prime}}^{\left(  -2\right)  }\right\rangle
\\
&  +\sum_{l^{\prime}}\omega_{ll^{\prime}}^{\left(  2\right)  }\left\langle
f_{l}^{\left(  0\right)  }-f_{l^{\prime}}^{\left(  0\right)  }\right\rangle
+\sum_{l^{\prime}}\omega_{ll^{\prime}}^{\left(  3\right)  }\left\langle
f_{l}^{\left(  -1\right)  }-f_{l^{\prime}}^{\left(  -1\right)  }\right\rangle
\\
&  +\sum_{l^{\prime}}\left[  \omega_{ll^{\prime}}^{\left(  4\right)
}+S_{ll^{\prime}}^{\left(  4\right)  }\right]  \left\langle f_{l}^{\left(
-2\right)  }-f_{l^{\prime}}^{\left(  -2\right)  }\right\rangle +\frac
{1}{i\hbar}C_{l}^{\prime\prime},
\end{align*}
where the expression \cite{KL1957,Luttinger1958} for $S_{ll^{\prime}}^{\left(
4\right)  }=S_{l^{\prime}l}^{\left(  4\right)  }$ is not necessary here. The
symmetric parts of the higher-order scattering rates only contribute to
trivial renormalizations to the longitudinal transport, and thus are
negligible in the case of weak disorder potential. $\frac{1}{i\hbar}%
C_{l}^{\left(  0\right)  }=\frac{1}{\hbar}e\mathbf{E}\cdot\partial
_{\mathbf{k}}\rho_{l}^{\left(  0\right)  }$ is the driving term of the
conventional Boltzmann equation, where $\rho_{l}^{\left(  0\right)  }$ is just
the Fermi distribution function \cite{KL1957}. $C_{l}^{\prime\prime}$ contains
the combination effects of the electric field and disorder, whose original
expression (is of the second order of disorder potential) is given in Ref.
\cite{Luttinger1958}. Here the point is that, except the anomalous part
$C_{l}^{\prime\prime,a}=-i\hbar e\mathbf{E\cdot}\sum_{l^{\prime}}^{\prime
}\omega_{ll^{\prime}}^{\left(  2\right)  }\delta\mathbf{\tilde{r}}_{l^{\prime
}l}\partial_{\epsilon_{l}}\rho_{l}^{\left(  0\right)  }$, all other terms of
$C_{l}^{\prime\prime}$ are present even in the absence of SOI and are trivial
renormalizations to the conventional driving term. Therefore, regarding the
spin/anomalous Hall effect, the qualitatively and quantitatively important
contributions to $\left\langle f_{l}\right\rangle $ in the weak
disorder-potential regime are given by
\begin{gather*}
0=\left[  e\mathbf{E}\cdot\mathbf{v}_{l}^{0}\partial_{\epsilon_{l}}\rho
_{l}^{\left(  0\right)  }+\sum_{l^{\prime}}\omega_{ll^{\prime}}^{\left(
2\right)  }\left\langle f_{l}^{\left(  -2\right)  }-f_{l^{\prime}}^{\left(
-2\right)  }\right\rangle \right] \\
+\left[  \sum_{l^{\prime}}\omega_{ll^{\prime}}^{\left(  2\right)
}\left\langle f_{l}^{\left(  -1\right)  }-f_{l^{\prime}}^{\left(  -1\right)
}\right\rangle +\sum_{l^{\prime}}\omega_{ll^{\prime}}^{3a}\left\langle
f_{l}^{\left(  -2\right)  }-f_{l^{\prime}}^{\left(  -2\right)  }\right\rangle
\right] \\
+\left[  \sum_{l^{\prime}}\omega_{ll^{\prime}}^{\left(  2\right)
}\left\langle f_{l}^{\left(  0\right)  ,n}-f_{l^{\prime}}^{\left(  0\right)
,n}\right\rangle +\sum_{l^{\prime}}\omega_{ll^{\prime}}^{3a}\left\langle
f_{l}^{\left(  -1\right)  }-f_{l^{\prime}}^{\left(  -1\right)  }\right\rangle
\right. \\
\left.  +\sum_{l^{\prime}}\omega_{ll^{\prime}}^{4a}\left\langle f_{l}^{\left(
-2\right)  }-f_{l^{\prime}}^{\left(  -2\right)  }\right\rangle \right] \\
+\left[  \sum_{l^{\prime}}\omega_{ll^{\prime}}^{\left(  2\right)
}\left\langle f_{l}^{\left(  0\right)  ,a}-f_{l^{\prime}}^{\left(  0\right)
,a}\right\rangle -e\mathbf{E\cdot}\sum_{l^{\prime}}\omega_{ll^{\prime}%
}^{\left(  2\right)  }\delta\mathbf{\tilde{r}}_{l^{\prime}l}\partial
_{\epsilon_{l}}\rho_{l}^{\left(  0\right)  }\right]  .
\end{gather*}
Here we split $f_{l}^{\left(  0\right)  }$ into $f_{l}^{\left(  0\right)
}=f_{l}^{\left(  0\right)  ,n}+f_{l}^{\left(  0\right)  ,a}$. Setting
$\left\langle f_{l}^{\left(  -2\right)  }\right\rangle \equiv g_{l}^{\left(
-2\right)  }$, $\left\langle f_{l}^{\left(  -1\right)  }\right\rangle \equiv
g_{l}^{\left(  -1\right)  }$, $\left\langle f_{l}^{\left(  0\right)
,n}\right\rangle \equiv g_{l}^{\left(  0\right)  }$ and $\left\langle
f_{l}^{\left(  0\right)  ,a}\right\rangle \equiv g_{l}^{a}$, the above
equation is just the form of the SB equations (\ref{SBE-n},\ref{SBE-a}) used
in practice \cite{Sinitsyn2008,Nagaosa2010}.

Because the Kohn-Luttinger expansion is basically a bare perturbation theory,
some trivial renormalization terms are unavoidable in high orders of this
expansion. These terms should be eliminated systematically by a
renormalization procedure, if one aims at placing the Kohn-Luttinger theory as
a generic foundation for the extended-state transport phenomena. This kind of
renormalization treatment has been shown for free electrons without SOI
\cite{Moore1967}. Although a more complicated procedure is expected to be
applicable also in the presence of SOI, it has never been done according to
our literature knowledge. In fact, the Kubo linear response approach may be
more suitable to serve as the foundation of the extended-state transport, into
which the systematic renormalization procedure can be incorporated. On the
other hand, we only regard the Kohn-Luttinger approach as a foundation of the
SB theory in the case of weak disorder potential. In this case the
aforementioned renormalization effects are just much smaller high-order
corrections to the longitudinal transport, and can thus be neglected in the
Boltzmann regime.

In the weak disorder-potential regime, after disorder average, one has%
\begin{gather}
\sum_{ll^{\prime}}^{\prime}\left\langle f_{ll^{\prime}}A_{l^{\prime}l}%
^{eq}\right\rangle =\sum_{ll^{\prime}}^{\prime}\left\langle f_{ll^{\prime}%
}^{\left(  0\right)  }\right\rangle \left\langle A_{l^{\prime}l}%
^{eq}\right\rangle +\sum_{ll^{\prime}}^{\prime}\left\langle f_{ll^{\prime}%
}^{\left(  -1\right)  }A_{l^{\prime}l}^{eq}\right\rangle \nonumber\\
=\sum_{ll^{\prime}}^{\prime}C_{ll^{\prime}}^{\left(  0\right)  }%
\frac{\left\langle A_{l^{\prime}l}^{eq}\right\rangle }{d_{ll^{\prime}}^{-}%
}\nonumber\\
+\sum_{ll^{\prime}l^{\prime\prime}}^{\prime}\left\langle W_{ll^{\prime\prime}%
}W_{l^{\prime\prime}l^{\prime}}\right\rangle \left\langle \frac{f_{l}^{\left(
-2\right)  }-f_{l^{\prime\prime}}^{\left(  -2\right)  }}{d_{ll^{\prime\prime}%
}^{-}}-\frac{f_{l^{\prime\prime}}^{\left(  -2\right)  }-f_{l^{\prime}%
}^{\left(  -2\right)  }}{d_{l^{\prime\prime}l^{\prime}}^{-}}\right\rangle
\frac{\left\langle A_{l^{\prime}l}^{eq}\right\rangle }{d_{ll^{\prime}}^{-}%
}\nonumber\\
+2\operatorname{Re}\sum_{ll^{\prime}}^{\prime}\left\langle f_{l}^{\left(
-2\right)  }\right\rangle \left\langle \frac{W_{l^{\prime}l}A_{ll^{\prime}%
}^{eq}}{d_{ll^{\prime}}^{+}}\right\rangle
\end{gather}
Due to $C_{ll^{\prime}}^{\left(  0\right)  }=ie\mathbf{E\cdot\tilde{J}%
}_{ll^{\prime}}\left(  \rho_{l^{\prime}}^{\left(  0\right)  }-\rho
_{l}^{\left(  0\right)  }\right)  $ and $\mathbf{v}_{ll^{\prime}}^{phy}%
\delta_{\mathbf{kk}^{\prime}}=-\frac{1}{i\hbar}\left(  \epsilon_{l}%
-\epsilon_{l^{\prime}}\right)  i\mathbf{\tilde{J}}_{ll^{\prime}}$ $\left(
l\neq l^{\prime}\right)  $, we have%
\begin{equation}
\sum_{ll^{\prime}}^{\prime}C_{ll^{\prime}}^{\left(  0\right)  }\frac
{\left\langle A_{l^{\prime}l}^{eq}\right\rangle }{d_{ll^{\prime}}^{-}}%
=\sum_{l}f_{l}^{0}\delta^{in}A_{l},
\end{equation}%
\begin{align}
&  \sum_{ll^{\prime}l^{\prime\prime}}^{\prime}\left\langle W_{ll^{\prime
\prime}}W_{l^{\prime\prime}l^{\prime}}\right\rangle \left\langle \frac
{f_{l}^{\left(  -2\right)  }-f_{l^{\prime\prime}}^{\left(  -2\right)  }%
}{d_{ll^{\prime\prime}}^{-}}-\frac{f_{l^{\prime\prime}}^{\left(  -2\right)
}-f_{l^{\prime}}^{\left(  -2\right)  }}{d_{l^{\prime\prime}l^{\prime}}^{-}%
}\right\rangle \frac{\left\langle A_{l^{\prime}l}^{eq}\right\rangle
}{d_{ll^{\prime}}^{-}}\nonumber\\
&  \equiv\sum_{l}\left\langle f_{l}^{\left(  -2\right)  }\right\rangle
\delta^{sj}A_{l},
\end{align}
and ($\hat{A}^{eq}=\hat{A}^{phy}+\delta^{V}\hat{A}$)%
\begin{equation}
2\operatorname{Re}\sum_{ll^{\prime}}^{\prime}\left\langle f_{l}^{\left(
-2\right)  }\right\rangle \left\langle \frac{W_{l^{\prime}l}A_{ll^{\prime}%
}^{eq}}{d_{ll^{\prime}}^{+}}\right\rangle =\sum_{l}\left\langle f_{l}^{\left(
-2\right)  }\right\rangle \delta^{sj,1}A_{l},
\end{equation}
with $\delta^{in}A_{l}$, $\delta^{sj}A_{l}$ and $\delta^{sj,1}A_{l}$ given by
Sec. III.\

Besides, $\sum_{l}\rho_{l}^{\left(  0\right)  }\left(  \delta^{\mathbf{E}}%
\hat{A}\right)  _{ll}=\sum_{l}f_{l}^{0}\delta^{in,1}A_{l}$, thus%
\begin{align}
&  \sum_{ll^{\prime}}^{\prime}\left\langle f_{ll^{\prime}}A_{l^{\prime}l}%
^{eq}\right\rangle +\sum_{l}\rho_{l}^{0}\left(  \delta^{\mathbf{E}}\hat
{A}\right)  _{ll}\\
&  =\sum_{l}f_{l}^{0}\left(  \delta^{in}A_{l}+\delta^{in,1}A_{l}\right)
+\sum_{l}g_{l}^{\left(  -2\right)  }\left(  \delta^{sj}A_{l}+\delta
^{sj,1}A_{l}\right)  .\nonumber
\end{align}

\section{Proof of Eqs. (\ref{sj-velocity}) and (\ref{Berry-velocity})}

\subsection{Proof of Eq. (\ref{sj-velocity})}

Following the same route shown in Appendix A of Ref. \cite{Xiao2017SOT-SBE},
we get%
\begin{align}
\delta^{sj}\mathbf{v}_{l}  &  =\operatorname{Re}\sum_{l^{\prime}}\frac
{2}{\hbar}\frac{\left\langle W_{ll^{\prime}}\left[  W,\mathbf{\tilde{J}%
}\right]  _{l^{\prime}l}\right\rangle _{c}}{d_{ll^{\prime}}^{-}}\nonumber\\
&  +\sum_{l^{\prime}}\frac{2\pi}{\hbar}\left\langle \left\vert W_{ll^{\prime}%
}\right\vert ^{2}\right\rangle _{c}\delta\left(  d_{ll^{\prime}}\right)
\left[  i\mathbf{\tilde{J}}_{l^{\prime}}-i\mathbf{\tilde{J}}_{l}\right]
\end{align}
with $\left[  W,\mathbf{\tilde{J}}\right]  _{l^{\prime}l}\equiv\sum
_{l^{\prime\prime}}\left[  W_{l^{\prime}l^{\prime\prime}}\mathbf{\tilde{J}%
}_{l^{\prime\prime}l}-\mathbf{\tilde{J}}_{l^{\prime}l^{\prime\prime}%
}W_{l^{\prime\prime}l}\right]  $. In the present case we have%
\begin{equation}
\left[  W,\mathbf{\tilde{J}}\right]  _{l^{\prime}l}=-i\left[  \hat
{W},\mathbf{\hat{r}}^{phy}\right]  _{l^{\prime}l}+\mathbf{\hat{D}}%
W_{l^{\prime}l},
\end{equation}
then%
\begin{align*}
\delta^{sj}\mathbf{v}_{l}  &  =\operatorname{Re}\sum_{l^{\prime}}\frac
{2}{\hbar}\frac{\left\langle W_{ll^{\prime}}\mathbf{\hat{D}}W_{l^{\prime}%
l}\right\rangle _{c}}{d_{ll^{\prime}}^{-}}\\
&  +\sum_{l^{\prime}}\frac{2\pi}{\hbar}\left\langle \left\vert W_{ll^{\prime}%
}\right\vert ^{2}\right\rangle _{c}\delta\left(  d_{ll^{\prime}}\right)
\left[  i\mathbf{\tilde{J}}_{l^{\prime}}-i\mathbf{\tilde{J}}_{l}\right] \\
&  -\operatorname{Re}\sum_{l^{\prime}}2\frac{\left\langle W_{ll^{\prime}%
}\left(  \frac{1}{i\hbar}\left[  \mathbf{\hat{r}}^{phy},\hat{W}\right]
\right)  _{l^{\prime}l}\right\rangle _{c}}{d_{ll^{\prime}}^{-}}.
\end{align*}
The first term on the rhs can be split into two terms with one related to
$\operatorname{Im}\left\langle W_{ll^{\prime}}\mathbf{\hat{D}}W_{l^{\prime}%
l}\right\rangle _{c}=\left\langle \left\vert W_{ll^{\prime}}\right\vert
^{2}\right\rangle _{c}\mathbf{\hat{D}}\arg W_{l^{\prime}l}$ and the other
related to $\mathbf{\hat{D}}\left\langle \left\vert W_{ll^{\prime}}\right\vert
^{2}\right\rangle _{c}$. The first one is related to the phase of the disorder
potential and is thus nontrivial. While the latter one, which does not break
any symmetry, is just a trivial renormalization to $\mathbf{v}_{l}^{0}$. It
does not contribute to the Hall current in the weak disorder-potential limit
and can be ignored. Thus one has%
\begin{align}
\delta^{sj}\mathbf{v}_{l}  &  =\sum_{l^{\prime}}\frac{2\pi}{\hbar}\left\langle
\left\vert W_{ll^{\prime}}\right\vert ^{2}\right\rangle _{c}\delta\left(
d_{ll^{\prime}}\right)  \left[  i\mathbf{\tilde{J}}_{l^{\prime}}%
-i\mathbf{\tilde{J}}_{l}-\mathbf{\hat{D}}\arg W_{l^{\prime}l}\right]
\nonumber\\
&  -\operatorname{Re}\sum_{l^{\prime}}2\frac{\left\langle W_{ll^{\prime}%
}\left(  \frac{1}{i\hbar}\left[  \mathbf{\hat{r}}^{phy},\hat{W}\right]
\right)  _{l^{\prime}l}\right\rangle _{c}}{d_{ll^{\prime}}^{-}}%
\end{align}

On the other hand%
\begin{equation}
\delta^{sj,1}\mathbf{v}_{l}=2\operatorname{Re}\sum_{l^{\prime}\neq l}%
\frac{\left\langle \left(  \frac{1}{i\hbar}\left[  \mathbf{\hat{r}}^{phy}%
,\hat{W}\right]  \right)  _{ll^{\prime}}W_{l^{\prime}l}\right\rangle _{c}%
}{d_{ll^{\prime}}^{+}},
\end{equation}
thus $\delta^{sj}\mathbf{v}_{l}+\delta^{sj,1}\mathbf{v}_{l}=\sum_{l^{\prime}%
}\omega_{ll^{\prime}}^{\left(  2\right)  }\delta\mathbf{\tilde{r}}_{l^{\prime
}l}$.

\subsection{Proof of Eq. (\ref{Berry-velocity})}

We choose the external in-plane electric field in x direction, then%
\begin{gather}
\left(  \delta^{in}v_{l}\right)  _{y}=-\hbar eE_{x}\sum_{l^{\prime}\neq
l}\delta_{\mathbf{kk}^{\prime}}\frac{2\mathrm{Im}\langle u_{l}|\hat{v}%
_{x}^{phy}|u_{l^{\prime}}\rangle\langle u_{l^{\prime}}|\hat{v}_{y}^{phy}%
|u_{l}\rangle}{d_{ll^{\prime}}^{2}}\nonumber\\
=-\frac{e}{\hbar}E_{x}\sum_{\eta^{\prime}\neq\eta}2\mathrm{Im}\left[  \langle
u_{\mathbf{k}}^{\eta}|\left(  i\partial_{k_{x}}+\hat{x}^{a}\right)
|u_{\mathbf{k}}^{\eta^{\prime}}\rangle\right. \nonumber\\
\left.  \langle u_{\mathbf{k}}^{\eta^{\prime}}|\left(  i\partial_{k_{y}}%
+\hat{y}^{a}\right)  |u_{\mathbf{k}}^{\eta}\rangle\right] \\
=\frac{e}{\hbar}E_{x}2\mathrm{Im}\langle\partial_{k_{y}}u_{\mathbf{k}}^{\eta
}|\partial_{k_{x}}u_{\mathbf{k}}^{\eta}\rangle-\frac{e}{\hbar}E_{x}%
2\mathrm{Im}\langle u_{\mathbf{k}}^{\eta}|\hat{x}^{a}\hat{y}^{a}%
|u_{\mathbf{k}}^{\eta}\rangle\nonumber\\
-\frac{e}{\hbar}E_{x}2\mathrm{\operatorname{Re}}\left[  \langle u_{\mathbf{k}%
}^{\eta}|\hat{x}^{a}|\partial_{k_{y}}u_{\mathbf{k}}^{\eta}\rangle
-\langle\partial_{k_{x}}u_{\mathbf{k}}^{\eta}|\hat{y}^{a}|u_{\mathbf{k}}%
^{\eta}\rangle\right] \nonumber\\
=\frac{e}{\hbar}E_{x}2\mathrm{Im}\langle\partial_{k_{y}}u_{\mathbf{k}}^{\eta
}|\partial_{k_{x}}u_{\mathbf{k}}^{\eta}\rangle\nonumber\\
+\frac{e}{\hbar}E_{x}\left[  \partial_{k_{x}}\langle u_{\mathbf{k}}^{\eta
}|\hat{y}^{a}|u_{\mathbf{k}}^{\eta}\rangle-\partial_{k_{y}}\langle
u_{\mathbf{k}}^{\eta}|\hat{x}^{a}\mathbf{|}u_{\mathbf{k}}^{\eta}\rangle\right]
\nonumber\\
-\frac{e}{\hbar}E_{x}\langle u_{\mathbf{k}}^{\eta}|\left(  \partial_{k_{x}%
}\hat{y}^{a}-\partial_{k_{y}}\hat{x}^{a}\right)  \mathbf{|}u_{\mathbf{k}%
}^{\eta}\rangle\nonumber\\
+i\frac{e}{\hbar}E_{x}\langle u_{\mathbf{k}}^{\eta}|\left[  \hat{x}^{a}%
,\hat{y}^{a}\right]  |u_{\mathbf{k}}^{\eta}\rangle\nonumber
\end{gather}
On the other hand, $\left(  \delta^{in,1}v_{l}\right)  _{y}=\left(
\delta^{\mathbf{E}}\hat{v}_{y}\right)  _{ll}$ where%
\begin{align}
\delta^{\mathbf{E}}\hat{v}_{y}  &  =\frac{1}{i\hbar}\left[  \hat{y}%
^{phy},-eE_{x}\hat{x}^{phy}\right] \\
&  =\frac{eE_{x}}{\hbar}\left(  \frac{\partial\hat{y}^{a}}{\partial\hat{k}%
_{x}}-\frac{\partial\hat{x}^{a}}{\partial\hat{k}_{y}}\right)  -i\frac{eE_{x}%
}{\hbar}\left[  \hat{x}^{a},\hat{y}^{a}\right]  ,\nonumber
\end{align}
thus%
\begin{gather}
\left(  \delta^{in}v_{l}\right)  _{y}+\left(  \delta^{in,1}v_{l}\right)
_{y}=\frac{eE_{x}}{\hbar}\left\{  2\mathrm{Im}\langle\partial_{k_{y}}%
u_{l}|\partial_{k_{x}}u_{l}\rangle+\right. \nonumber\\
\left.  \partial_{k_{x}}\langle u_{\mathbf{k}}^{\eta}|\hat{y}^{a}%
|u_{\mathbf{k}}^{\eta}\rangle-\partial_{k_{y}}\langle u_{\mathbf{k}}^{\eta
}|\hat{x}^{a}\mathbf{|}u_{\mathbf{k}}^{\eta}\rangle\right\} \nonumber\\
=\frac{eE_{x}}{\hbar}\left(  \Omega_{l}^{0}+\Omega_{l}^{a}\right)  .
\end{gather}

\section{Kubo-Streda formula in effective quantum theories}

In the linear response Eq. (\ref{linear-response}) $\delta A=tr\left\langle
\hat{A}^{eq}\hat{f}\right\rangle +tr\left\langle \hat{\rho}\delta^{\mathbf{E}%
}\hat{A}\right\rangle $ to the dc uniform external electric field, the first
term on the rhs can be found by the Kubo-Streda formula
\cite{Streda2010,Bruno2001} in terms of the $\hat{A}-\mathbf{\hat{\jmath}}$
correlation function, in the presence of static impurities. Starting from the
single-particle quantum Liouville equation, one has \cite{Streda2010}
\begin{equation}
\hat{f}=\frac{i}{\hbar}\lim_{s\rightarrow0^{+}}\int_{-\infty}^{0}%
dte^{st}e^{i\hat{H}^{eq}t/\hbar}\left[  \hat{\rho},\hat{H}_{1}\right]
e^{-i\hat{H}^{eq}t/\hbar}, \label{Kubo}%
\end{equation}
where $\hat{H}^{eq}$ is the equilibrium single-carrier Hamiltonian in the
presence of disorder. When the physical position operator is just the
canonical one, $\hat{H}^{eq}=\hat{H}_{0}+\hat{V}\left(  \mathbf{\hat{r}%
}\right)  $ and $\hat{H}_{1}=-e\mathbf{E\cdot\hat{r}}$, then Eq. (\ref{Kubo})
leads to the well-known Kubo-Streda formula for the $\hat{A}-\mathbf{\hat
{\jmath}}$ correlation function \cite{Nagaosa2010,Sinova2015}. In this case
$\delta^{\mathbf{E}}\hat{A}=0$ when $\hat{A}$ is not the thermal current
operator, thus $\delta A=tr\left\langle \hat{A}^{eq}\hat{f}\right\rangle $ is
completely determined by the $\hat{A}-\mathbf{\hat{\jmath}}$ correlation function.

In effective quantum theories where the physical position operator is not the
canonical one, $\hat{H}^{eq}=\hat{H}_{0}+\hat{V}\left(  \mathbf{\hat{r}}%
^{phy}\right)  $, $\hat{H}_{1}=-e\mathbf{E\cdot\hat{r}}^{phy}$ and
$\delta^{\mathbf{E}}\hat{A}\neq0$ even when $\hat{A}$ is the electric/spin
current operator. In this case we derive the slightly generalized Kubo-Streda
formula for $\delta A$, by obtaining first the Bastin formula by the method in
Ref. \cite{Streda2010} and then following the manipulations presented in Ref.
\cite{Bruno2001}. At the low-temperature limit the generalized Kubo-Streda
formula for the spin Hall conductivity reads $\sigma_{yx}^{z}=\sigma
_{yx}^{z,I\left(  a\right)  }+\sigma_{yx}^{z,I\left(  b\right)  }+\sigma
_{yx}^{z,II}+\sigma_{yx}^{z,in-1}$, where
\begin{equation}
\sigma_{yx}^{z,I\left(  a\right)  }=\frac{\hbar}{2\pi}tr\left\langle
\hat{\jmath}_{y}^{z,eq}\hat{G}^{R}\left(  \epsilon_{F}\right)  \hat{\jmath
}_{x}^{eq}\hat{G}^{A}\left(  \epsilon_{F}\right)  \right\rangle ,
\label{surface-a}%
\end{equation}%
\begin{equation}
\sigma_{yx}^{z,I\left(  b\right)  }=-\frac{\hbar}{2\pi}\operatorname{Re}%
tr\left\langle \hat{\jmath}_{y}^{z,eq}\hat{G}^{R}\left(  \epsilon_{F}\right)
\hat{\jmath}_{x}^{eq}\hat{G}^{R}\left(  \epsilon_{F}\right)  \right\rangle ,
\label{surface-b}%
\end{equation}%
\begin{gather}
\sigma_{yx}^{z,II}=\frac{\hbar}{2\pi}\operatorname{Re}\int d\epsilon
f^{0}\left(  \epsilon\right) \label{sea}\\
\times tr\left\langle \hat{\jmath}_{y}^{z,eq}\hat{G}^{R}\left(  \epsilon
\right)  \hat{\jmath}_{x}^{eq}\frac{d\hat{G}^{R}\left(  \epsilon\right)
}{d\epsilon}-\hat{\jmath}_{y}^{z,eq}\frac{d\hat{G}^{R}\left(  \epsilon\right)
}{d\epsilon}\hat{\jmath}_{x}^{eq}\hat{G}^{R}\left(  \epsilon\right)
\right\rangle ,\nonumber
\end{gather}
and $\sigma_{yx}^{z,in-1}=\sum_{l}f_{l}^{0}\langle l|\frac{1}{2}\left\{
\frac{\hbar}{2}\hat{\sigma}_{z},\delta^{\mathbf{E}}\hat{v}_{y}\right\}
\mathbf{|}l\rangle/E_{x}$, i.e.,%
\begin{equation}
\sigma_{yx}^{z,in-1}=\sum_{l}f_{l}^{0}\langle l|\frac{1}{2}\left\{
\frac{\hbar}{2}\hat{\sigma}_{z},\frac{-e}{i\hbar}\left[  \hat{y}^{phy},\hat
{x}^{phy}\right]  \right\}  \mathbf{|}l\rangle. \label{additional}%
\end{equation}
Here%
\begin{equation}
\mathbf{\hat{\jmath}}^{eq}=\frac{e}{i\hbar}\left[  \mathbf{\hat{r}}^{phy}%
,\hat{H}_{0}+\hat{W}\right]  ,\mathbf{\hat{\jmath}}^{z,eq}=\frac{1}{2}\left\{
\frac{\hbar}{2e}\hat{\sigma}_{z},\mathbf{\hat{\jmath}}^{eq}\right\}  ,
\label{Kubo-effective}%
\end{equation}
and $\hat{G}^{R/A}\left(  \epsilon\right)  =\left(  \epsilon-\hat{H}^{eq}\pm
i0^{+}\right)  ^{-1}$ is the retarded/advanced Green's function operator. In
Eq. (\ref{additional}) only the disorder-free part of $\sigma_{yx}^{z,in-1}$
is retained, because we focus on the weak disorder-potential regime. In this
regime $\sigma_{yx}^{z,I\left(  b\right)  }$ can be neglected and in
$\sigma_{yx}^{z,II}$ only the disorder-free part is important
\cite{Nagaosa2010,Sinova2015}. Applying Eqs. (\ref{surface-a}) -
(\ref{additional}) to the 2D Nozieres-Lewiner effective model, we get
$\sigma_{yx}^{z,II}=0$, $\sigma_{yx}^{z,I\left(  a\right)  }=\sigma
_{yx}^{z,sj}+\sigma_{yx}^{z,ad}$, $\sigma_{yx}^{z,sj}=\sigma_{yx}%
^{z,ad}=-\sigma_{yx}^{z,in-1}$ and $\sigma_{yx}^{z}=\sigma_{yx}^{z,ad}$. For
the Rashba 2D effective model, one can also obtain the same result as the SB approach.

On the other hand, in usual applications of the Kubo-Streda formula to spin
Hall effects in the 2D Nozieres-Lewiner effective model
\cite{Bruno2001,Dugaev2001,Tse2006PRL,Tse2006PRB}, $\hat{H}^{eq}=\hat{H}%
_{0}+\hat{W}$ but $\hat{H}_{1}=-e\mathbf{E\cdot\hat{r}}$, and the physical
position is not distinguished from the canonical position. Then those authors
used $\sigma_{yx}^{z,0}=\sigma_{yx}^{z,I\left(  a\right)  ,0}+\sigma
_{yx}^{z,I\left(  b\right)  ,0}+\sigma_{yx}^{z,II,0}$ where%
\[
\sigma_{yx}^{z,I\left(  a\right)  ,0}=\frac{\hbar}{2\pi}tr\left\langle
\hat{\jmath}_{y}^{z,eq,0}\hat{G}^{R}\left(  \epsilon_{F}\right)  \hat{\jmath
}_{x}^{eq,0}\hat{G}^{A}\left(  \epsilon_{F}\right)  \right\rangle ,
\]%
\[
\sigma_{yx}^{z,I\left(  b\right)  ,0}=-\frac{\hbar}{2\pi}\operatorname{Re}%
tr\left\langle \hat{\jmath}_{y}^{z,eq,0}\hat{G}^{R}\left(  \epsilon
_{F}\right)  \hat{\jmath}_{x}^{eq,0}\hat{G}^{R}\left(  \epsilon_{F}\right)
\right\rangle ,
\]%
\begin{gather*}
\sigma_{yx}^{z,II,0}=\frac{\hbar}{2\pi}\operatorname{Re}\int d\epsilon
f^{0}\left(  \epsilon\right)  tr\left\langle \hat{\jmath}_{y}^{z,eq,0}\hat
{G}^{R}\left(  \epsilon\right)  \hat{\jmath}_{x}^{eq,0}\frac{d\hat{G}%
^{R}\left(  \epsilon\right)  }{d\epsilon}\right. \\
\left.  -\hat{\jmath}_{y}^{z,eq,0}\frac{d\hat{G}^{R}\left(  \epsilon\right)
}{d\epsilon}\hat{\jmath}_{x}^{eq,0}\hat{G}^{R}\left(  \epsilon\right)
\right\rangle ,
\end{gather*}
with%
\begin{equation}
\mathbf{\hat{\jmath}}^{eq,0}\equiv\frac{e}{i\hbar}\left[  \mathbf{\hat{r}%
},\hat{H}_{0}+\hat{W}\right]  ,\mathbf{\hat{\jmath}}^{z,eq,0}\equiv\frac{1}%
{2}\left\{  \frac{\hbar}{2e}\hat{\sigma}_{z},\mathbf{\hat{\jmath}}%
^{eq,0}\right\}  . \label{Kubo-usual}%
\end{equation}
Applying this usual Kubo-Streda formula to the 2D case of $\hat{H}_{0}%
=\frac{\hbar^{2}\mathbf{\hat{k}}^{2}}{2m}$, one has $\sigma_{yx}^{z,II,0}=0$,
$\sigma_{yx}^{z,I\left(  a\right)  ,0}=\sigma_{yx}^{z,0,sj}+\sigma
_{yx}^{z,0,ad}$ and $\sigma_{yx}^{z,0,sj}=\sigma_{yx}^{z,0,ad}=\frac{1}%
{2}\sigma_{yx}^{z,0,ad}$, thus $\sigma_{yx}^{z,0}=\sigma_{yx}^{z,ad}$.

\section{Technical details in Sec. VI}

For 2D electrons, $\mathbf{\hat{r}}^{a}=\frac{\lambda_{0}^{2}}{4}\hat{\sigma
}_{z}\mathbf{\hat{z}\times\hat{k}}$ and $i\mathbf{J}_{l}^{a}=\frac{\lambda
_{0}^{2}}{4}\left(  \hat{\sigma}_{z}\right)  _{ll}\mathbf{\hat{z}\times k}$.
Thus $\left(  \delta^{in,1}\mathbf{v}_{l}\right)  _{y}=\frac{eE_{x}}{\hbar
}2\frac{\lambda_{0}^{2}}{4}\left(  \hat{\sigma}_{z}\right)  _{ll}$ and
$\sum_{l}f_{l}^{0}\left(  \delta^{in,1}\mathbf{v}_{l}\right)  _{y}%
=\frac{eE_{x}}{\hbar}2\frac{\lambda_{0}^{2}}{4}\sum_{l}f_{l}^{0}\left(
\hat{\sigma}_{z}\right)  _{ll}$, meanwhile
\begin{align}
&  \sum_{l}g_{l}^{\left(  -2\right)  }\delta^{sj,1}\mathbf{v}_{l}%
\label{Vignale-1}\\
&  =2\sum_{ll^{\prime}}g_{l}^{\left(  -2\right)  }\omega_{l^{\prime}%
l}^{\left(  2\right)  }\frac{\lambda_{0}^{2}}{4}\operatorname{Re}\left(
\hat{\sigma}_{z}\right)  _{ll^{\prime}}\mathbf{\hat{z}}\times\left(
\mathbf{k}^{\prime}-\mathbf{k}\right) \nonumber\\
&  -4\frac{\lambda_{0}^{2}}{4\hbar}\sum_{ll^{\prime}}^{\prime}g_{l}^{\left(
-2\right)  }\frac{\left\langle \left\vert V_{l^{\prime}l}\right\vert
^{2}\right\rangle }{d_{ll^{\prime}}}\operatorname{Im}\left(  \hat{\sigma}%
_{z}\right)  _{ll^{\prime}}\mathbf{\hat{z}}\times\left(  \mathbf{k}^{\prime
}-\mathbf{k}\right)  .\nonumber
\end{align}
In the 2D Nozieres-Lewiner model $\left(  \hat{\sigma}_{z}\right)
_{ll^{\prime}}=\eta\delta_{\eta\eta^{\prime}}$, thus
\begin{align}
&  \sum_{l}g_{l}^{\left(  -2\right)  }\left(  \delta^{sj,1}\mathbf{v}%
_{l}\right)  _{y}\label{Vignale-2}\\
&  =2\sum_{\mathbf{k}^{\prime},\mathbf{k}\eta}g_{\eta\mathbf{k}}^{\left(
-2\right)  }\omega_{\eta\mathbf{k}^{\prime},\eta\mathbf{k}}^{\left(  2\right)
}i\left[  \mathbf{J}_{\eta\mathbf{k}^{\prime}}^{a}-\mathbf{J}_{\eta\mathbf{k}%
}^{a}\right]  _{y}\nonumber\\
&  =2\frac{e}{\hbar}E_{x}\sum_{\eta\mathbf{k}}\partial_{k_{x}}f^{0}\left(
i\mathbf{J}_{\eta\mathbf{k}}^{a}\right)  _{y}=-\frac{e}{\hbar}E_{x}\sum
_{\eta\mathbf{k}}f_{\eta\mathbf{k}}^{0}\Omega_{\eta\mathbf{k}}^{a}\nonumber\\
&  =-\sum_{l}f_{l}^{0}\left(  \delta^{in,1}\mathbf{v}_{l}\right)
_{y}.\nonumber
\end{align}
Here $\left(  \delta^{in,1}\mathbf{v}_{l}\right)  _{y}=\frac{eE_{x}}{\hbar
}\Omega_{\eta\mathbf{k}}^{a}$ is only valid for the considered model (Appendix
B). But if $\left\langle u_{l}\left\vert \hat{\sigma}_{z}\right\vert
u_{l^{\prime}}\right\rangle $ has $l$-off-diagonal matrix elements or is
momentum dependent, we cannot get the relation $\sum_{l}g_{l}^{\left(
2\right)  }\left(  \delta^{sj,1}\mathbf{v}_{l}\right)  _{y}=-\sum_{l}f_{l}%
^{0}\left(  \delta^{\mathbf{E}}\mathbf{v}\right)  _{l}$.

Next we analyze the case of the conventional spin current in 2D hole systems.
By $\mathbf{\hat{r}}^{a}=\frac{\lambda_{0}^{2}}{4}\hat{\sigma}_{z}%
\mathbf{\hat{z}\times\hat{K}}$,
\begin{equation}
\sum_{l}f_{l}^{0}\left(  \delta^{in,1}\mathbf{j}_{l}^{z}\right)  _{y}%
=E_{x}\frac{e\lambda_{0}^{2}}{4}\sum_{l}f_{l}^{0}\frac{\frac{\partial K_{x}%
}{\partial k_{x}}+\frac{\partial K_{y}}{\partial k_{y}}}{2},
\end{equation}
and $\hat{V}\left(  \mathbf{\hat{r}}^{phy}\right)  =\hat{V}\left(
\mathbf{\hat{r}}\right)  +\frac{1}{2}\left(  \mathbf{\nabla}\hat{V}%
\cdot\mathbf{\hat{r}}^{a}+\mathbf{\hat{r}}^{a}\cdot\mathbf{\nabla}\hat
{V}\right)  $ leads to
\begin{gather}
\left(  \delta^{V}\mathbf{\hat{\jmath}}_{y}^{z}\right)  _{ll^{\prime}}%
=\frac{1}{2i}\frac{\lambda_{0}^{2}}{4}V_{ll^{\prime}}\left(  K_{x}%
-K_{x}^{\prime}\right) \nonumber\\
+\frac{i}{4}\frac{\lambda_{0}^{2}}{4}V_{ll^{\prime}}\left(  \left(
\mathbf{k}-\mathbf{k}^{\prime}\right)  \times\mathbf{\hat{z}}\right)
\cdot\left(  \frac{\partial\mathbf{K}^{\prime}}{\partial k_{y}^{\prime}}%
+\frac{\partial\mathbf{K}}{\partial k_{y}}\right)  ,
\end{gather}
then%
\begin{gather}
\sum_{l}g_{l}^{\left(  -2\right)  }\left(  \delta^{sj,1}\mathbf{j}_{l}%
^{z}\right)  _{y}=\frac{\hbar}{2}\frac{\lambda_{0}^{2}}{4}\sum_{ll^{\prime}%
}g_{l}^{\left(  -2\right)  }\omega_{l^{\prime}l}^{\left(  2\right)  }\left[
\left(  K_{x}^{\prime}-K_{x}\right)  \right. \nonumber\\
\left.  +\left(  k_{x}^{\prime}-k_{x}\right)  \frac{1}{2}\left(
\frac{\partial K_{x}^{\prime}}{\partial k_{x}^{\prime}}+\frac{\partial K_{x}%
}{\partial k_{x}}\right)  \right. \nonumber\\
\left.  -\left(  k_{y}^{\prime}-k_{y}\right)  \frac{1}{2}\left(
\frac{\partial K_{x}^{\prime}}{\partial k_{y}^{\prime}}+\frac{\partial K_{x}%
}{\partial k_{y}}\right)  \right]  , \label{Vignale-3}%
\end{gather}
where we have used $\frac{\partial K_{x}}{\partial k_{x}}=\frac{\partial
K_{y}}{\partial k_{y}}$ for 2D electrons and 2D holes. For 2D electrons,
$\mathbf{\hat{K}=\hat{k}}$\ thus the last term in the square brackets on the
rhs of Eq. (\ref{Vignale-3}) vanishes and we obtain the previous result. While
for 2D holes, $\mathbf{\hat{K}}=\left(  \hat{k}_{x}^{3}-3\hat{k}_{y}^{2}%
\hat{k}_{x},3\hat{k}_{x}^{2}\hat{k}_{y}-\hat{k}_{y}^{3},0\right)  $ and
$\sum_{l}g_{l}^{\left(  -2\right)  }\left(  \delta^{sj,1}\mathbf{j}_{l}%
^{z}\right)  _{y}$ is unlikely to be equal to $\sum_{l}f_{l}^{0}\left(
\delta^{in,1}\mathbf{j}_{l}^{z}\right)  _{y}$ in general cases. Even if for
slowly varying disorder potentials, the minus sign of the last term in the
square brackets on the rhs of Eq. (\ref{Vignale-3}) makes further
simplification impossible, unless some special assumptions are made for the
model Hamiltonian.

\bigskip

\bigskip

\bigskip

\begin{thebibliography}{99}
\bibitem {note-regime}The Kubo-Streda linear response formula is valid from
the Boltzmann regime to the diffusive regime.
\end{thebibliography}
\end{document}